\def\presentation{
\voffset -.50in  \hoffset -.19in
\oddsidemargin 0in \evensidemargin 0in
\marginparwidth .75in \marginparsep 7pt \topmargin 0in
\headheight 12pt \headsep .25in
\footheight 18pt \footskip .35in
\textheight 9.5in \textwidth 6.5in
\columnsep 10pt \columnseprule 0pt }
\documentstyle{article}
\presentation
\begin{document}
%

%
\def\tilde{\widetilde}
\def\bar{\overline}
\def\hat{\widehat}
\def\*{\star}
\def\[{\left[}
\def\]{\right]}
\def\({\left(}
\def\){\right)}
\def\zb{{\bar{z} }}
\def\frac#1#2{{#1 \over #2}}
\def\inv#1{{1 \over #1}}
\def\half{{1 \over 2}}
\def\d{\partial}
\def\der#1{{\partial \over \partial #1}}
\def\dd#1#2{{\partial #1 \over \partial #2}}
\def\vev#1{\langle #1 \rangle}
\def\bra#1{{\langle #1 |  }}
\def\ket#1{ | #1 \rangle}
\def\rvac{\hbox{$\vert 0\rangle$}}
\def\lvac{\hbox{$\langle 0 \vert $}}
\def\2pi{\hbox{$2\pi i$}}
\def\e#1{{\rm e}^{^{\textstyle #1}}}
\def\grad#1{\,\nabla\!_{{#1}}\,}
\def\dsl{\raise.15ex\hbox{/}\kern-.57em\partial}
\def\Dsl{\,\raise.15ex\hbox{/}\mkern-.13.5mu D}
\def\comm#1#2{ \BBL\ #1\ ,\ #2 \BBR }
\def\x{\stackrel{\otimes}{,}}
\def\det{ {\rm det}}
\def\tr{{\rm tr}}
\def\bacs{\backslash}
%
\def\th{\theta}		\def\Th{\Theta}
\def\ga{\gamma}		\def\Ga{\Gamma}
\def\be{\beta}
\def\al{\alpha}
\def\ep{\epsilon}
\def\la{\lambda}	\def\La{\Lambda}
\def\de{\delta}		\def\De{\Delta}
\def\om{\omega}		\def\Om{\Omega}
\def\sig{\sigma}	\def\Sig{\Sigma}
\def\vphi{\varphi}
%
%
\def\CA{{\cal A}}	\def\CB{{\cal B}}	\def\CC{{\cal C}}
\def\CD{{\cal D}}	\def\CE{{\cal E}}	\def\CF{{\cal F}}
\def\CG{{\cal G}}	\def\CH{{\cal H}}	\def\CI{{\cal J}}
\def\CJ{{\cal J}}	\def\CK{{\cal K}}	\def\CL{{\cal L}}
\def\CM{{\cal M}}	\def\CN{{\cal N}}	\def\CO{{\cal O}}
\def\CP{{\cal P}}	\def\CQ{{\cal Q}}	\def\CR{{\cal R}}
\def\CS{{\cal S}}	\def\CT{{\cal T}}	\def\CU{{\cal U}}
\def\CV{{\cal V}}	\def\CW{{\cal W}}	\def\CX{{\cal X}}
\def\CY{{\cal Y}}	\def\CZ{{\cal Z}}
%
%
\catcode`\@=11
\font\tenmsa=msam10 at 12truept
\font\sevenmsa=msam7
\font\fivemsa=msam5
\font\tenmsb=msbm10 at 12truept
\font\sevenmsb=msbm7 at 9truept
\font\fivemsb=msbm5 at 7truept
\newfam\msafam
\newfam\msbfam
\textfont\msafam=\tenmsa \scriptfont\msafam=\sevenmsa
 \scriptscriptfont\msafam=\fivemsa
\textfont\msbfam=\tenmsb \scriptfont\msbfam=\sevenmsb
 \scriptscriptfont\msbfam=\fivemsb
 
\def\hexnumber@#1{\ifcase#1 0\or1\or2\or3\or4\or5\or6\or7\or8\or9\or
        A\or B\or C\or D\or E\or F\fi }
 
\def\msb{\tenmsb\fam\msbfam}
\def\Bbb{\ifmmode\let\next\Bbb@\else
 \def\next{\errmessage{Use \string\Bbb\space only in math mode}}\fi\next}
\def\Bbb@#1{{\Bbb@@{#1}}}
\def\Bbb@@#1{\fam\msbfam#1}
\def\Cmath{\Bbb C}
\def\Rmath{\Bbb R}
\def\Zmath{\Bbb Z}
%
\def\cadremath#1{\vbox{\hrule\hbox{\vrule\kern8pt\vbox{\kern8pt
			\hbox{$\displaystyle #1$}\kern8pt} 
			\kern8pt\vrule}\hrule}}
\def\proof{\noindent Proof. \hfill \break}
\def\cqfd{ {\hfill{$\Box$}} }
\def\square{\hfill
\vrule height6pt width6pt depth1pt \\}
%
%
\def\debut{ \begin{eqnarray} }
\def\fin{ \end{eqnarray} }
\def\non{ \nonumber }
%

%
%
\rightline{SPhT-97/113}
\rightline{LPT ENS-97/41}
\vskip 1cm
\centerline{\LARGE Twisted Self-Duality of Dimensionally Reduced Gravity}
\bigskip
\centerline{\LARGE and Vertex Operators.}
\vskip 2cm
\centerline{\large  D. Bernard~${}^a$ ${}^1$, 
B. Julia~${}^b$ \footnote[1]{Member of the CNRS} }
\bigskip
\centerline{${}^a$~~~Service de Physique Th\'eorique de Saclay
\footnote[2]{\it Laboratoire de la Direction des Sciences de la
Mati\`ere du Commissariat \`a l'Energie Atomique.}}
\centerline{F-91191, Gif-sur-Yvette, France.}
\centerline{ ${}^b$~~~Laboratoire de physique th\'eorique CNRS-ENS}
\centerline{24 rue Lhomond, F-75231 Paris Cedex 05, France.}
\vskip2cm
Abstract. \footnote[3]{  Research supported in part by EC under TMR
contract ERBFMRX-CT96-0012 \vskip -12pt} \\
The Geroch group, isomorphic to the SL(2,R) affine Kac-Moody group,  
is an infinite dimensional solution generating group of Einstein's equations 
with two surface orthogonal commuting Killing vectors. We introduce 
another solution generating group for these equations, the dressing 
group, and discuss its connection with the Geroch group. We 
show that it acts transitively on a dense subset of moduli
space. We use a new Lax pair 
expressing a twisted self-duality of this system and we study the 
dressing problem associated to it. We also describe how to use 
vertex operators to solve the reduced Einstein's equations. In particular 
this allows to find solutions by purely algebraic computations.

%
%
%
\newpage
%
%
\def\bacs{\backslash}
\section{Introduction.}

The quantization of gravitation is an outstanding problem 
that will require inductive jumps and where the correspondence 
principle is largely insufficient.  A possible 
attitude is to study invariant subspaces of the full phase space and to try
to quantize these systems. Einstein's equations do exhibit chaotic aspects as 
well as integrable ones, 
here we shall be concerned with the latter but one should keep
in mind that noncompact simple Lie groups entail negative curvature and chaos. 
It has been known for 25 years that 2D reduced gravity with two commuting
Killing vectors admits a large solution generating
 group \cite{ger}, the so-called 
Geroch group, and for 19 years that it is an integrable classical
theory \cite{zaka,mai1}. The conjecture that the 
Geroch group, suitably redefined to keep track of the central extension,
is isomorphic to the $SL(2,\Rmath)$ affine Kac-Moody group  (or $A^{(1)}_1$,
which in this paper will always mean the split real form, namely the real form
obtained from the Weyl basis)  was first made
in \cite{julia} where the Chevalley-Serre generators have been identified.
The Geroch group acts on the space of configurations of an infinite set of 
(dual) potentials covering the space of
metrics of General Relativity. The analytical difficulties come
from the choice of normalisation condition for the definition of these 
auxiliary fields, and that is already delicate locally.
The connection between the integrable nature of the
system, or more precisely the existence of a (generalised) Lax pair,
 and the existence of this group was clarified in \cite{mais}. Formally 
the structure is quite close to that of flat space sigma models for
symmetric spaces yet there is no direct correspondence between the two 
problems. As a matter of fact it will turn out that the gravity model will
resemble more the conformal affine Toda systems than the flat space sigma 
models.  Another important guideline was to keep a unified view of all 
supergravities, and this allowed a systematic study of hidden symmetries of 
gravity itself as the bosonic sector of N=1 supergravity. 
Progress was prompted by their 
simultaneous analysis in 3 dimensions \cite{giha, creju}.

The integrability of 2D reduced gravity relies on the existence of
an associated linear system whose compatibility implies 
Einstein's equations.  However, the algebraic structure underlying 
this system was obscured until recently by 
the presence of so-called moving poles, due to the spacetime dependence
of the scattering parameter of the linear system of \cite{mais}. It is our
choice to define spacetime derivatives for constant scattering parameter or not
but two different choices differ by derivatives with respect to the scattering 
parameter, a possibility that was not considered before.
We first avoid this difficulty by introducing a new Lax connexion
without any moving poles which is a slight generalization of the
one introduced to the same effect in \cite{julanico}. This is achieved 
once more by making use of the conformal internal symmetry which extends the 
Geroch group. This allows us to bridge the remaining gap between the
standard methods of integrable field theories, including the method of 
dressing transformations \cite{zakasha, zaka, semen} and the group theoretical
solution generating procedures cherished by relativists \cite{kinn} and by
high energy physicists. A third 
point of view is provided by the twistor methods see \cite{mato}.
Yet another approach uses isomonodromy see eg. \cite{nico}.

We construct the group of dressing transformations and  then
compare it to the Geroch group. They are both solution
generating groups. We also show that (in a local sense) most of phase space 
is made of one and only one dressing
orbit. As usual the elements of the dressing group are pairs of elements 
$(g_-,g_+)$
where $g_\pm$ belong respectively to two opposite  Borel subgroups of 
$A^{(1)}_1$. The latter will appear as the   
twisted affine Kac-Moody group (which could be slightly abusively called 
$A^{(2)}_1$ \cite{lewi}) 
but is isomorphic to $A^{(1)}_1$ in the so-called 
principal gradation for which both simple roots are given equal weights. In 
order to distinguish it from the dressing group we shall call it the father 
group! Strictly speaking, prior to the socalled Weyl coordinate 
choice, the dressing involves internal conformal transformations as well.
The multiplication law of the dressing group is:
\debut
(g_-,g_+)\cdot (h_-,h_+)= (g_-h_-,g_+h_+) \label{mul}
\fin
The dressing group can be endowed with a Poisson structure 
so that it turns into a Lie-Poisson group \cite{semen}.
Comparing with very similar integrable systems, it is tempting
to conjecture that these dressing transformations 
are actually Lie-Poisson actions on the phase space and that
after quantization they will turn into quantum group symmetries.
Note that the twist uses an involution of the finite dimensional symmetry
algebra already present in 3 dimensions.

In fact our discussion will at first apply to general symmetric space sigma 
models reduced from curved three-dimensional space to 2 dimensions. 
But then we shall make use 
of the particular symmetry of pure gravity (reduced from 4 to 2 dimensions) 
to simplify the construction of solutions. Namely
there exist two dual formulations of 2D reduced gravity.
Let us parametrize the metric by fields $(\hat \sig, \De,  N)$ as:
\debut
ds^2 = 2 \rho^{1/2} e^{2\hat \sig}~ da db + \rho\De^{-1}dx^2
+ \rho \De (dy - N dx)^2 \label{001}
\fin
where $\rho = a+b$ has been hidden in a coordinate choice. Note that it is 
generically possible (and  only locally) to  use as coordinates $z^\nu$
the chiral parts $a$ and $b$ of the scalar harmonic field $\rho= a+b$.
As a result, and admittedly only locally, the internal conformal
symmetry mentioned previously becomes the ordinary spacetime conformal group.
Now, given any solution of the  reduced Einstein's equations and in any 
coordinates, we get a
dual solution by introducing the metric $ds^{* 2}$ parametrized
as above, eq.(\ref{001}), but with the original fields
$(\hat \sig, \De, N)$ replaced by
the dual fields $(\hat \sig^*, \De^*, N^*)$ with:
\debut
\De^* = 1/(\rho\De) \quad,\quad
\De^*\d_\nu N^*= \ep_{\nu \al} \De \d_\al N 
\quad,\quad \De^* e^{4\hat \sig^*} = \De e^{4 \hat \sig} \label{dual0}
\fin
$\rho$ remains inert under the duality.
The  duality we just defined is the Kramer-Neugebauer 
symmetry exchanging the two 
$SL(2)$ subgroups generating the Geroch group. 
For more general curved space (with symmetric target space)
sigma models in 3 dimensions reduced to 2D one may presumably use 
further symmetries of the affine Dynkin diagram when they exist.

We introduce $\tau$-functions for 2D reduced (classical) gravity. 
They may be defined as expectation values of 
elements of the father group, $A^{(1)}_1$, in auxiliary Fock spaces
which serve as representation spaces of the group. 
They encode all local informations about a given
solution and are directly related to the original fields
$(\hat \sig, \De, N)$ or their derivatives.
For example the conformal factor $\hat \sig$ may be written as
a product of two $\tau$-functions:
\debut
e^{2 \hat \sig} = 
\bra{\La_+} \Psi_0\cdot (g_-^{-1}g_+)\cdot \Psi_0^{-1} \ket{\La_+}\cdot
\bra{\La_-} \Psi_0\cdot (g_-^{-1}g_+)\cdot \Psi_0^{-1} \ket{\La_-}
\label{002}
\fin
where $\ket{\La_\pm}$ are specific vacuum states in Fock space representations 
of the father dressing algebra. 
$\Psi_0$ is the so-called vacuum wave function which carries all the
coordinate dependence, it takes its values in a 2 dimensional solvable subgroup 
of the Moebius subgroup of the conformal group. $g_\pm$ are  a pair of
operators defining one element 
of the dressing group and parametrizing a  solution. It would be instructive
to compare this formula and the formula for the conformal factor obtained in 
\cite{koni}.
We also describe how to use vertex operators to compute these
$\tau$-functions. Vertex operators provide tools to algebraically 
solve Riemann-Hilbert problems.

The dual formulation leads to a very compact algebraic formula for the metric. 
Indeed, as we are going to explain, 
all the dual fields $\hat \sig^*$, $\De^*$ and $N^*$
are computable by evaluating some matrix elements in the
auxiliary Fock spaces. In particular,
\debut
N^* \pm \frac{i}{\De^*}&=& 
\frac{\bra{\La_\pm} \Psi_0\cdot (g_-^{-1}\hat \CE^*g_+)\cdot \Psi_0^{-1} \ket{\La_\pm}}{
\bra{\La_\pm} \Psi_0\cdot (g_-^{-1}g_+)\cdot \Psi_0^{-1} \ket{\La_\pm}}
\label{003}
\fin
where  $\ket{\La_\pm}$,  $\Psi_0$ and $g_\pm$ are as in eq.(\ref{002}),
$\hat \CE^*$ represents the same Lie algebra element on the two   Fock representation spaces; it
turns out that the two representatives differ by a sign.
All the expectation values in
(\ref{002},\ref{003}) are computable using vertex operators.
This provides a way to determine metrics that are
 solutions of Einstein's equations
without any integration beyond the decomposition (once and for all) of  $\rho$
according to $\rho =a+b$ . 
The formula (\ref{003}) for the dual variables stems from the
fact (discussed in subsection 5.3) that the dual fields $\De^*,~N^*$ can be used as coordinates in 
a two dimensional vector subspace of the dressing algebra generated  
by conjugating $\hat \CE^*$ by the solution of the auxiliary linear system.
\bigskip

The paper is organized as follows. 
In section 2, we introduce a few basic facts 
concerning Einstein's equations with two commuting
Killing vectors and their relations with sigma models.
In section 3, we introduce the new Lax connection for this system 
of equations and present its basic properties. 
In section 4, we define the dressing group and
we describe its connection with the Geroch group.
In section 5, we introduce the $\tau$-functions and prove
the algebraic formulas (\ref{001},\ref{002},\ref{003}) for the metric.
In section 6, we explain how to use vertex operators to evaluate
the necessary expectation values.
In section 7, we give explicit examples of metrics solutions of Einstein's
equations and illustrate the general methods of the paper.
Five appendices summarize technical informations.
 
\section{Einstein's equations and $H\bacs G$ models.}
In this section we review a few well known facts concerning Einstein's
equations with two commuting Killing vectors and their relations with
$H\bacs G$ sigma models. Cf. eg. \cite{nico} for a review.

\subsection{Einstein's equations.}
Assuming that there are two commuting Killing vectors, orthogonal
\footnote{This is no further restriction for asymptotically flat solutions} 
to a foliation by surfaces, one looks for metrics of the form \cite{zaka}:
\debut
ds^2 &=& f(\sig,t) (-dt^2 + d\sig^2) + \rho(\sig,t) S_{ij}(\sig,t) dx^idx^j 
\label{metric}\\
 &=&2f(z_+,z_-)dz_-dz_+  + \rho(z_-,z_+) S_{ij}(z_-,z_+) dx^idx^j 
\non
\fin
where $z_\pm$ are the light-cone coordinates: $z_\pm= (\sig\pm t)/\sqrt{2}$.
The symmetric $2\times 2$ matrix $S_{ij}$ is normalized by $det[S]=1$.
Thus $S$ is an element of $SL(2,\Rmath)$.
Clearly this parametrization is covariant under conformal
transformations.
We shall assume for definiteness that the two Killing vectors are space-like 
and we shall start from 4 dimensional Minkowski signature (unless stated 
otherwise).

Introducing the internal  zweibein $\sqrt{\rho}\CV_j^a$, we have:
\debut
\rho~S_{ij} dx^idx^j = \rho~(dx^i \CV_i^a)\de_{ab}(\CV_j^bdx^j) \non
\fin
In matrix form: $S= {}^t\CV\cdot \CV$, where ${}^t\CV$ denotes the transposition of $\CV$. 
The matrix $\CV$ also belongs to $SL(2,\Rmath)$.

The local $SO(2)$ gauge symmetry, which acts by 
$\CV^a_j(z_\pm)\to O^a_b(z_\pm)\cdot \CV^b_j(z_\pm)$
with $O(z_\pm)\in SO(2)$, is manifest. 
There is also a global $SL(2,\Rmath)$ symmetry which acts by $\CV^a_j(z_\pm)\to
\CV^a_k(z_\pm) M^k_j$ with $M\in SL(2,\Rmath)$. It corresponds to a global redefinition
of the coordinates $x^k\to M^{-1}{}^k_jx^j$.
For future convenience we recall our conventions: the local $SO(2)$
transformations act on $\CV$ on the left and the global $SL(2,\Rmath)$ acts on
the right:
\debut
SO(2)_{local} \longrightarrow \CV(z_+,z_-) \longleftarrow SL(2,\Rmath)_{global} \non
\fin

Einstein's equations, which demand the metric (\ref{metric}) to be
Ricci flat, are then equivalent to the following set of equations \cite{zaka}:
\debut
\d_+\d_-\rho &=& 0 \label{E1}\\
\d_+(\rho \d_-S.S^{-1}) + \d_-(\rho \d_+S.S^{-1}) &=& 0 \label{E2}\\
(\rho^{-1}\d_\pm\rho)\cdot (\d_\pm\hat \sig) &=& -\inv{8} tr(\d_\pm S \d_\pm S^{-1}) 
\label{E3}
\fin 
Here and below $\d_\pm=\d_{z_\pm}$, and $\hat \sig = \log\la -
\half \log(\d_+\rho\d_-\rho)$ and $\la^2=\rho^{-\half} f$.

Eq. (\ref{E1}) means that $\rho$ is harmonic, so it decomposes as:
\debut
\rho(z_+,z_-)= a(z_+) + b(z_-) \label{aetb}
\fin
Using the conformal covariance of the parametrization (\ref{metric}),
one may generically
 choose $a(z_+)$ and $b(z_-)$ as the two light-cone coordinates.  
The metrics (\ref{metric}) may then be written as:
\debut
ds^2 = 2 \rho^{\half} e^{2\hat \sig} da(z_+)db(z_-) + \rho S_{ij} dx^idx^j 
\label{metric2}
\fin
Eq. (\ref{E2}) is a non-linear equation for $S$; it expresses the conservation 
of the current $(\rho \d_\pm S.S^{-1})$ corresponding to the right invariance 
under $SL(2,\Rmath)_{global}$. Eq. (\ref{E3}) determines $\hat \sig$
once $\rho$ and $S$ are known. It determines $\hat \sig$ up to an
additive constant, which up to a rescaling of the coordinates $x^j$
corresponds to a global rescaling of the metric $ds^2$.
Note that Minkowski or chiral spaces are not generic, they 
can only be recovered as limits by the
dressing method. So it is conceptually important to preserve
coordinate reparametrization invariance \cite{julanico}, in practice however 
coordinates may  be chosen most appropriately.

\subsection{$H\bacs G$ models.}

The reduced Einstein equations (\ref{E1},\ref{E2},\ref{E3}) can be written 
in a gauge covariant way by formulating them as a sigma model on
the coset space $SO(2)\bacs SL(2,\Rmath)$.
It is equally simple to consider the general case of sigma models
on $H\bacs G$ where $G$ is a non-compact group and $H$ the maximal
compact subgroup. Let $\bf g$ and $\bf h$ be the Lie
algebra of $G$ and $H$ respectively; $\bf h$ is a Lie
subalgebra of $\bf g$. There exists an order two
automorphism $\eta$ of $\bf g$ such that $\bf h$ is
its eigenvalue one subspace, i.e. $\eta(\bf h) =\bf h$ pointwise.
This automorphism can be extended to the group $G$ such
that elements of $H$ are $\eta$-invariant.
The Lie algebra $\bf g$ decomposes as :
\debut
{\bf g} = {\bf h} + {\bf {r}} \quad {\rm with} \quad \eta(\bf h) =\bf h, 
\quad \eta(\bf r) =-\bf r \non
\fin
with $\bf r$ a representation of $\bf h$ and
 $[\bf h,\bf h]\subset \bf h$, $[\bf h,\bf r]\subset \bf r$ and 
$[\bf r,\bf r]\subset \bf h$.

The simplest examples correspond to $G=SL(n,\Rmath)$ and $H=SO(n)$, for which
the automorphism is $\eta(g)= {}^t g^{-1}$. At the Lie algebra
level, this corresponds to $\eta(X)=-{}^t X$.
Thus elements of $\bf h$ are antisymmetric matrices and those of
$\bf r$ are symmetric traceless matrices.
For $n=2$, $H=SO(2)$ is abelian, we choose the following basis
for the Lie algebra: $\bf h =\{ \sig^+-\sig^- \}$ 
and $\bf r = \{ \sig^z~,~ \sig^++\sig^- \}$ with
$\sig^a$ Pauli matrices
\footnote{We use the conventions: $[\sig^z,\sig^\pm]=\pm 2\sig^\pm$ and
$[\sig^+,\sig^-]=\sig^z$, and $tr(\sig^z\sig^z)=2tr(\sig^+\sig^-)=2$} .

To rewrite Einstein's equations one introduces the G-invariant and H-covariant
connection  $\d \CV\cdot \CV^{-1}\in \bf g$
and denote its components on $\bf h$ and $\bf r$ by $-Q$ and $-P$,
respectively:
\debut
\d_{\pm}\CV\cdot \CV^{-1}= -Q_\pm - P_\pm, \quad {\rm with}\quad
Q_\pm\in {\bf h},~~ P_\pm \in {\bf r} 
\fin
The equations for $Q$ and $P$ are then, cf eg.\cite{nico}: 
\debut
\d_+Q_--\d_-Q_+ + [Q_+,Q_-] + [P_+,P_-] &=&0 \label{EL1}\\
D_-P_+-D_+P_-=0 \label{EL2}\\
D_-(\rho P_+) + D_+(\rho P_-) =0 \label{EL3}\\
(\d_\pm\hat \sig)(\rho^{-1}\d_\pm\rho)= \half tr(P_\pm P_\pm) \label{EL4}
\fin
where $D_\pm = \d_\pm + [Q_\pm,~\cdot~]$ is the covariant derivative. 
The first two equations mean that $(Q_\pm+P_\pm)$ is flat;
this ensures that there exists $\CV\in G$ such that 
$\d_{\pm}\CV\cdot \CV^{-1}= -Q_\pm - P_\pm$. 
Given the connexion $(Q_\pm+P_\pm)$, the $n$-bein $\CV$ is determined by 
quadratures. It is easy to verify that eqs.(\ref{EL1}-\ref{EL4}) imply
eqs.(\ref{E1}-\ref{E3}) for $S=\eta(\CV)^{-1}\cdot \CV$, since
they imply $S^{-1}\d_\pm S=-2\CV^{-1}P_\pm \CV$.
Taking the derivative of eq.(\ref{EL4}) implies:
\debut
\d_+\d_- \hat \sig = -\half tr(P_-P_+) \label{EL5}
\fin
This system of equations is gauge covariant. Gauge transformations
leave $\rho$ and $\hat \sig$ invariant and act on $Q_\pm$
and $P_\pm$ as:
\debut
Q_\pm &\to& Q^\La_\pm= \La Q_\pm \La^{-1} + \La \d_\pm \La^{-1} \label{gauge} \\
P_\pm &\to& P^\La_\pm=\La P_\pm \La^{-1}, \qquad~~~~~~~~~~~~~
{\rm with}\quad\La\in H \non
\fin
Physically inequivalent field configurations correspond to the data
of the field $\hat \sig$ up to translations (ie. constant shifts), 
and of the field
$\CV$ up to right multiplications by constant elements in $G$ and up to
gauge transformations by elements in $H$ acting on the left.
In other words, the space of inequivalent fields is parametrized by
the connexion $(Q_\pm+P_\pm)$ up to gauge transformations and
by $\hat \sig$ up to translations. We recall that $\rho$ has been hidden  
as a coordinate.

\subsection{The triangular gauge.}
The pure gravity case corresponds to $H\bacs G=SO(2)\bacs SL(2,\Rmath)$.
It is often convenient to choose a gauge in which $\CV$ is triangular:
\debut
\CV= \pmatrix{\De^{-\half} & 0 \cr -N \De^\half & \De^\half \cr}
=\pmatrix{\De^{-\half} & 0 \cr 0& \De^\half \cr}
\pmatrix{1 & 0 \cr -N  & 1 \cr}
\label{Etri}
\fin
The corresponding metric is : 
\debut
ds^2=2\rho^{\half} e^{2\hat \sig} dadb + \rho\De^{-1} dx^2
+\rho\De(dy - Ndx)^2 \label{metric3}
\fin
The connexion $(Q_\pm+P_\pm)$ is then 
\debut
Q_\pm+P_\pm =\half (\De^{-1}\d_\pm\De)\cdot \sig^z + \De(\d_\pm N)\cdot \sig^- 
\label{QPtri}
\fin
ie. $Q_\pm=-\half\De(\d_\pm N)\cdot(\sig^+-\sig^-)$ and
$P_\pm =\half (\De^{-1}\d_\pm\De)\cdot \sig^z +
\half\De(\d_\pm N)\cdot(\sig^++\sig^-)$.
The data of the flat connexion $(Q_\pm+P_\pm)$ in
the triangular gauge determines uniquely $\De$ and $N$ up to 
dilatations acting inversely on $\De$ and $N$ and up to translations
of $N$.  In this gauge the equations of motion (\ref{EL1}-\ref{EL3}) become:
\debut
\d_-\(\rho\De^2\d_+N\) + \d_+\(\rho\De^2\d_-N\) &=& 0 \label{Etri1}\\
\d_-\(\rho\De^{-1}\d_+\De\) + \d_+\(\rho\De^{-1}\d_-\De\)
&=& 2\rho\De^2(\d_+N)(\d_-N) \label{Etri2} \\
(\d_\pm\hat \sig)(\rho^{-1}\d_\pm\rho)&=& \inv{4}(\De^{-1}\d_\pm\De)^2
+\inv{4}(\De\d_\pm N)^2 \label{Etri3}
\fin
These are the so-called Ernst equations \cite{ernst}.
By construction these equations are $SL(2,\Rmath)$ invariant.
The $SL(2,\Rmath)$ action on $\CV$ is defined in the following way. 
Using the Iwasawa decomposition we can present $\CV$ as
$\CV= O\cdot kn$ where $O\in SO(2)$, $k$ is a diagonal matrix and $n$ is a 
lower triangular matrix with one on the diagonal.
The triangular gauge corresponds to choose $O$ equals the identity.
We can act on $\CV$ on the right by $SL(2,\Rmath)$ elements. 
Acting with diagonal or lower triangular matrices preserves
the triangular gauge, while acting on the right with an upper triangular
matrix breaks this gauge and one therefore has to compensate it
by a gauge transformation on the left.
The infinitesimal generators $J_z,~ J_\pm$ of these $SL(2,\Rmath)$ transformations
act as:
\debut
(J_z(\De), J_z(N), J_z(\hat \sig))&=&(\De, -N, 0 )  \label{ehl1}\\
(J_-(\De), J_-(N), J_-(\hat \sig))&=&(0, 1, 0 )  \label{ehl2}\\
(J_+(\De), J_+(N), J_+(\hat \sig))&=&(2\De N, -N^2+\De^{-2}, 0 )  \label{ehl3}
\fin
They satisfy the commutation relations:
$[J_z,J_\pm]=\pm J_\pm$ and $[J_+,J_-]=2J_z$.
By construction, $J_z$ and $J_-$ leave the connexion $(Q_\pm+P_\pm)$
invariant, wheras $J_+$ acts as an infinitesimal gauge transformation,
required for the preservation of the triangular gauge.
The corresponding gauge transformation is
$\La= 1 +\ep \De^{-1}(\sig^+-\sig^-)$:
\debut
J_z(Q_\pm+P_\pm)= 0\quad &;&\quad J_z(\d_\pm \hat \sig)= 0 \non\\
J_-(Q_\pm+P_\pm)= 0 \quad &;&\quad J_-(\d_\pm \hat \sig)= 0
\label{defmm}\\
J_+(Q_\pm+P_\pm)= (\d_\pm N)\cdot \sig^z+ 2(\d_\pm\De^{-1})\cdot \sig^-
\quad &;&\quad J_+(\d_\pm \hat \sig) = 0 \non
\fin
This $SL(2,\Rmath)$ group has been called the ``Matzner-Misner" group.
Finally recall the invariance under translation of $\hat \sig$.
Let $K$ be its generator:
\debut
(K(\De),K(N),K(\hat \sig))= (0,0,1/2) \label{defK}
\fin
It extends the $SL(2,\Rmath)$ group to a $GL(2,\Rmath)$ group.

\section{The Lax connection.}
In this section we introduce a new Lax connection for the 
eqs.(\ref{EL1}-\ref{EL3}) and(\ref{EL5} ) and we describe a few of its properties.
It is a modification in the central charge sector of the one proposed in \cite{julanico}.
This linear system only implies the second order equation for $\hat \sig$;
the first order equation will arise by imposing a supplementary
constraint on the wave function. This constraint is dictated in the 
older linear system by the choice of normalisation condition at the special value of the 
spectral parameter. 

\subsection{Algebraic notations.}
Let $\CH_{taf}$ be the affine Kac-Moody algebra over $\bf g$ 
twisted by the order two automorphism $\eta$ leaving $\bf h$ invariant. 
Let us beware that $\CH_{taf}$ is most definitely not the affine Kac-Moody 
algebra associated to ${\bf h}$ since it requires twisting. 
We denote its elements by $X\otimes t^n$ with
$n\in \Zmath$ and $X\in {\bf h}$ for $n$ even and $X\in {\bf r}$ for $n$ odd.
The central element is denoted by $k$. The commutation relations are:
\debut
\[{ X\otimes t^n, Y\otimes t^m }\]=
[X,Y]\otimes t^{n+m} +n \frac{k}{2} ~tr(XY)\de_{n+m,0} \label{comh}
\fin
We will similarly denote by $H_{taf}$ the twisted affine Kac-Moody 
group whose Lie algebra is $\CH_{taf}$; technically we shall consider formal power series in
$t$ or/and $t^{-1}$  . It will be the principal part 
of the father group.  

We can view $H_{taf}$ as a subgroup of the untwisted
affine Kac-Moody (formal) group $G_{af}$ whose
Lie algebra $\CG_{af}$ is generated by all the elements of the
form $X\otimes t^n$ with $n\in \Zmath$ and $X\in \bf g$ with no restriction.
The automorphism $\eta$ extends to an automorphism $\tau$ of $\CG_{af}$
with:
\debut
\tau(X\otimes t^n) = (-)^n\eta(X)\otimes t^n ,
\quad \tau(k)=k \label{invo}
\fin
The algebra $\CH_{taf}$ is the $\tau$-invariant subalgebra of $\CG_{af}$,
ie. $\tau(\CH_{taf})=\CH_{taf}$ pointwise. 
Remark that the bilinear form $(X\otimes t^n,Y\otimes t^m)=tr(XY)\de_{n+m,0}$
is neither positive nor negative definite on $\CH_{taf}$ and that it contains
the central generator. We may remark at this stage that the automorphism
$\tau$ is the same as in the work of Breitenlohner and Maison yet the change of 
parameter $s:= \frac{1-t}{1+t}$ changes the structure of its fixed point set at 
the formal level, in particular it now contains the Kac-Moody central generator.
For polynomials in the variable s, the fixed set was the maximal compact 
subalgebra as proposed in \cite{julia}, it did not contain the central 
generator.
 
Let $Vir$ be the Virasoro algebra. It is generated by elements
$L_n:= -\frac{1}{2} t^{2n+1} \partial_t$, $n\in \Zmath$,
with commutation relations:
\debut
\[{L_n,L_m}\]&=& (n-m) L_{n+m} + \frac{c}{12}n(n^2-1)\de_{n+m,0} \non
\fin
with $c$ the central element of the Virasoro algebra.
We need to introduce the elements
\debut
E_\pm = L_0-L_{\pm1} \label{defE}
\fin
of the Virasoro algebra. They satisfy the commutation relation of the traceless
2 by 2 triangular matrices ${\cal Tr}^2$:
$[E_+,E_-]= E_++ E_-$. The corresponding group we call $Tr^2$.
 The triangularity is manifest if one uses the other
variable $q = \frac{1-t^2}{1+t^2}$, $q$ appears in \cite{julanico}. 
This implies the reordering formula:
\debut
A^{-E_+}\cdot B^{-E_-} =
\({\frac{A}{A+B-1}}\)^{E_-}\cdot \({\frac{B}{A+B-1}}\)^{E_+} 
\label{reorder}
\fin
Its proof will be given in eqs.(\ref{psizero}) below, we shall also investigate 
its range of validity.

\def\bH{ {\bf VH}}
\def\bh{  V\CH}
Let $\bh$ be the semidirect product of the Virasoro algebra by  the
affine Kac-Moody algebra $\CH_{taf}$.  The crossed commutation relations are:
\debut
\[{L_n, X\otimes t^m}\] &=& -\frac{m}{2} X\otimes t^{m+2n} \label{virmo}
\fin
Note that the Virasoro generators $L_n$ shift the Kac-Moody degrees by $2n$.
In other words, the grading we use to count the degree is $2L_0$.
We will denote by $\bH$ an 
infinite dimensional group whose Lie algebra is $\bh$.
\bigskip 

We need to introduce a factorization problem in the affine algebra
$\CH_{taf}$, as well as in $H_{taf}$. It is an algebraic version
of a Riemann Hilbert problem \cite{semen}. Let $\CB_\pm$ be the two Borel
subalgebras of $\CH_{taf}$ respectively generated by elements
$(X\otimes t^{\pm n})$ with $n\geq 0$. Then any element $\hat X \in \CH_{taf}$
can be decomposed (almost uniquely) as:
\debut
\hat X = \hat X_+ - \hat X_-, \quad {\rm with} \quad
\hat X_\pm \in (\CB_\pm\oplus \Cmath k) \label{factor}
\fin
We shall demand that the components of $\hat X_\pm$ on the central element $k$
be opposite. Since the intersection between $\CB_\pm$ is the
algebra $\bf h$, the decomposition (\ref{factor}) is unique only
up to translations of $\hat X_\pm$ by elements of $\bf h$.
This freedom will correspond to our  gauge invariance.
But to be specific and to have uniqueness of the decomposition
(\ref{factor}) one may for example require that $\hat X_-$ has
no component on $\bf h$.

This factorization problem formally extends to the affine Kac-Moody group $H_{taf}$.
An element $g\in H_{taf}$ will be factorized as:
\debut
g = g_-^{-1}g_+, \quad {\rm with} \quad g_\pm \in \exp(\CB_\pm\oplus \Cmath k)
\label{factor2}
\fin
where we also demand that the components of $g_\pm$ on the exponential
of the central element be inverse. 
Once again we have the gauge freedom to multiply $g_\pm$ by
elements of $H$:
\debut
g_\pm \to h g_\pm \quad,\quad h\in H \label{ghpm}
\fin
To have uniqueness of the factorization,
one may for example  require that $g_-$ has no component on $H$.  Strictly
speaking the decomposition exists (at best) 
only generically again \cite{preseg}.
\bigskip

We will have to consider highest weight representations of $\bh$.
We shall require neither unitarizability nor reality of the representations.
These representations possess highest weight vectors $\ket{\La}$ satisfying:
$(X\otimes t^n)\ket{\La} = L_n \ket{\La} =0$ for $n>0$.   
Moreover $\ket{\La}$ is a highest weight vector for the finite   
dimensional algebra $\bf h$ generated by the elements $(X\otimes t^0)$,  
$X\in {\bf h}$. We will denote by $\ket{\la}$ the vectors of the
corresponding representation of $\bf h$, which we will call    
the highest representation of $\La$. They also satisfy
$(X\otimes t^n)\ket{\la} = L_n \ket{\la} =0$ for $n>0$. 
The vectors $\ket{\la}$ are simultaneously 
eigenvectors of $L_0$: $L_0\ket{\la}= h_\La \ket{\la}$;
$h_\La$ is called the conformal weight of $\La$.
The representations on the dual spaces are defined by $\vev{\hat X\cdot u,v}
=\vev{u,\om(\hat X)\cdot v}$ for $v$ in the representation 
$\La$ and $u$ in its dual, and $\om$ is the antihomomorphism defined by 
$\om (X\otimes t^n)={}^tX\otimes t^{-n}$.
Thus the dual representations are also highest weight representations.  
The dual vectors $\bra{\la}$  of the highest
representation of $\La$ satisfy
$\bra{\la}(X\otimes t^{-n})=\bra{\la} L_{-n}=0$ for $n>0$.
Notice that these defining properties imply that
$E_+\ket{\la}=h_\La\ket{\la}$   and $\bra{\la} E_-= \bra{\la}h_\La$.
\bigskip 

When writing explicit expressions for the metric we will 
restrict ourselves to the case of pure gravity. 
The real Lie algebra $\CH_{taf}$ is then isomorphic to the affine
algebra $A_1^{(1)}$ in the principal gradation.
We will need to consider complex representations
and abandon their unitarity.
The affine algebra $A_1^{(1)}$ possesses two
fundamental highest weight vectors $\ket{\La_\pm}$ which
are essentially characterized by:
\debut
(\sig^+-\sig^-)\ket{\La_\pm}&=&
\pm\frac{i}{2}\ket{\La_\pm},\quad {\rm and} \quad
k\ket{\La_\pm}=\ket{\La_\pm} \label{lapm}\\
(\sig^+-\sig^-)\otimes t^n\ket{\La_\pm}&=&
(\sig^++\sig^-)\otimes t^n\ket{\La_\pm}
=\sig^z\otimes t^n\ket{\La_\pm} = 0,
\quad {\rm for}\quad n>0 \non
\fin
The associated complex representations are constructed in terms of 
vertex operators acting on Fock spaces in Section 6.1.
\bigskip

\subsection{The Lax connection and the vacuum solution.}
We write the above equations of motion
(\ref{EL1},\ref{EL2},\ref{EL3}) and (\ref{EL5}) as a zero curvature condition
for a Lax connection. It is defined by:
\debut
A_\pm = \pm d_\pm E_\pm + Q_\pm + P_\pm\otimes t^{\pm1} 
\mp (\d_\pm\hat \sig) \frac{k}{2} 
\label{Lax}
\fin
Note that $A_\pm \in (\Cmath E_\pm \oplus \CB_\pm \oplus \Cmath k )$.
It is covariant under gauge transformations:
$A_\pm\to A_\pm^\La= \La A_\pm \La^{-1} + \La \d_\pm \La^{-1}$.
The zero curvature condition $[\d_++A_+,\d_-+A_-]=0$ is equivalent to:
\debut
d_\pm = \rho^{-1} \d_\pm \rho \quad {\rm with} \quad \d_-\d_+\rho=0 \label{L1}
\fin
together with the equations (\ref{EL1},\ref{EL2},\ref{EL3}) for
$Q_\pm$ and $P_\pm$, and the equation (\ref{EL5}) for $\hat \sig$.
\bigskip

The  matrix wave function $\Psi(z_\pm; \rho, \hat \sig, \De, X)   $ 
is defined to be the solution of the linear system,
\debut
(\d_\pm + A_\pm)~\Psi=0,\label{lin}
\fin
or equivalently,
\debut
(\d_\pm\Psi)\cdot \Psi^{-1} = -A_\pm  \quad \in {\bh}
\label{linear}
\fin
suitably normalized by some initial data.
Eq.(\ref{linear}) is a compatible differential system provided
the zero curvature condition is satisfied, ie. provided the
equations of motion are fulfilled.
Prior to normalization, this equation defines $\Psi$ up to 
multiplications on the right by constant elements. 
The wave function transforms covariantly under a gauge transformation
(\ref{gauge})~: $\Psi \to \Psi^\La=\La\cdot \Psi$ with $\La\in H$.
We also may extend the gauge algebra by $L_0$ and $k$. We shall call the 
resulting algebra the {\it local} gauge algebra to distinguish it from 
compensating {\it nonlocal} transformations necessary for the dressing action. 
The consequences of
the $L_0$ gauge transformation were discussed in \cite{julanico}.
Gauge transformation along $k$ modifies the coefficients of $\d \hat \sig$
in the Lax connection without changing the equations of motion.
The normalisation could be fixed by imposing the value of $\Psi$ at a 
given point, for instance at infinity. However in order to implement
the Geroch group there is a subtlety: 
one cannot restrict $\Psi$ to lie inside the semidirect product of $Tr^2$ or 
the M\" obius $SL(2,R)$ by $H_{taf}$ nor even to lie in $\bH$ as the form of 
the connection 
would permit but it will belong to the semidirect product of $\bf Vir$ by
$G_{af}$.
\bigskip

If one works in the semidirect product of $SL(2,R)$ by $H_{taf}$, 
the Lax connection satisfies the `universal' twisted self-duality
equation:
\debut
*~A^\perp =  \Om\cdot A^\perp \label{selfdual}
\fin
where $*$ is the Hodge operation, 
$A^\perp$ is the component of the connection orthogonal
to the local gauge Lie algebra and $\Om$ is an invariant operator
for that local gauge group of square one. The invariant bilinear form needed
for the orthogonal projection is such
that $L_0$ and $k$ are conjugate null vectors \cite{kac}.
In our case the local gauge Lie algebra is generated by $L_0, {\bf h}$
and $k$. So $A^\perp= \mp d_\pm L_{\pm 1} + P_\pm\otimes t^{\pm 1}$.
The invariant $\Om$ coincides with the derivation which
acts as $L_0$ on the M\"obius subalgebra and as $2L_0$ on the
Kac-Moody subalgebra. Eq.(\ref{selfdual}) selects in $A_\pm^\perp$
the elements of degree $\pm 1$.
\bigskip

An obvious solution of the equations of motion is
of course $Q_\pm=P_\pm=\hat \sig=0$. We will call it the vacuum 
solution. It depends on the dilaton field $\rho$ but locally and generically
the latter can be chosen as a coordinate. The affine dressing does not act on 
it so we can fix it once and for all in the present local discussion. We noted 
already that flat space for instance is nongeneric 
and that constant $\rho$ will lead to trivial dressing. 
 The Lax connection then reduces to 
$A_\pm= A^v_\pm$ where $A^v_\pm=\pm d_\pm ~E_\pm\in Vir $ with 
$d_\pm = \rho^{-1}\d_\pm\rho$ and $\rho=a(z_+)+b(z_-)$.
We write the wave function as $\psi_V$. It is solution
of the linear system in the Virasoro algebra,
\debut
(\d_\pm + A^v_\pm)~\psi_V=0 \label{laxvir}
\fin 
Two alternative expressions for $\psi_V$, which differ by the way
we order $E_\pm$, are:
\debut
\psi_V &=&
\({\frac{b(z_-)+c_1}{\rho}}\)^{E_+}~ \({\frac{b(z_-)+c_1}{c_2}}\)^{E_-}
\label{psiv1}\\
&=& \({\frac{\rho}{a(z_+)+c_3}}\)^{E_-}~ \({\frac{c_4}{a(z_+)+c_3}}\)^{E_+}
\label{psiv2}
\fin
where $c_1,~c_2,\cdots$ are constants.
The constants $c_3,~c_4$ are functions of $c_1$ and $c_2$; they depend
on the initial condition we choose. 
Let us sketch the proof of the formula (\ref{psiv1},\ref{psiv2}).
Decompose $\psi_V$ as $\psi_V= e^{\al E_+}\cdot e^{\beta E_-}$.
We have:
\debut
(\d_\pm\psi_V)\psi_V^{-1}&=&
(\d_\pm \al) E_+ + (\d_\pm\beta) e^{\al E_+}E_-e^{-\al E_+} \non\\
&=& [(\d_\pm \al) + (\d_\pm\beta)(e^\al-1)] E_+
+ e^\al (\d_\pm\beta) E_- \non
\fin
where we have used the formula:
$e^{\al E_+}E_-e^{-\al E_+}= e^\al E_- + (e^\al-1)E_+$.
The linear system $(\d_\pm\psi_V)\psi_V^{-1}= \mp d_\pm E_\pm$
then gives linear differential equations for $\al$ and
$\beta$ whose solutions are $\al= \log\({\frac{b(z_-)+c_1}{\rho}}\)$
and $\beta=\log\({\frac{b(z_-)+c_1}{c_2}}\)$.
This proves formula (\ref{psiv1}). The other formula is
proved in a similar way but using a different order for $E_\pm$.

It is convenient, but not necessary, to choose the initial condition
such that $\psi_V$ is equal to the identity at the point
where $a(z_+)=b(z_-)=1/2$. This requires $c_1=c_3=1/2$ and $c_2=c_4=1$.
We will denote by $\Psi_0$ this vacuum wave function:
\debut   
\Psi_0(z_+,z_-) &=&
\({\frac{b(z_-)+\inv{2}}{\rho}}\)^{E_+}~ \({b(z_-)+\inv{2}}\)^{E_-}~\non\\
&=& \({\frac{\rho}{a(z_+)+\inv{2}}}\)^{E_-}\({a(z_+)+\inv{2}}\)^{E_+}
\label{psizero}  
\fin
The reordering formula (\ref{reorder}) follows from the equality
between the two expressions for $\psi_V$
with this initial condition.

Another class of simple solutions is described in Appendix B.

\subsection{Projection onto the affine subalgebra and moving poles.}
Let us explain how we may recover the older Lax connection, with
its moving poles, from the one we just defined. The basic idea
is to first integrate the projection of the linear problem 
into the Virasoro subalgebra using the semidirect product structure of
the algebra $\bh$.

More precisely, let us decompose $A_\pm$ as:
$A_\pm= A^v_\pm+ {\hat A^{km}_\pm}$ where $A^v\in Vir$ and 
${\hat A^{km}}\in \CH_{taf}$;
ie. $A^v_\pm=\pm d_\pm ~E_\pm $ with $d_\pm =
\rho^{-1}\d_\pm\rho$ and $\rho=a(z_+)+b(z_-)$.
Write $\Psi= \hat \psi \cdot\psi_V$ where $\psi_V$ is solution
of the projection of the linear system into the Virasoro subalgebra, 
ie. $(\d_\pm + A^v_\pm)~\psi_V=0$. Expressions for $\psi_V$
have been given above in eqs.(\ref{psiv1},\ref{psiv2}). 
Then, $\hat \psi$ is solution of the linear problem:
\debut   
(\hat D_\pm  + {\hat A_\pm^{km}}~)~\hat \psi=0 \label{akmbis}
\fin 
where $\hat D_\pm$ is the covariant derivative defined by:
$\hat D_\pm \hat \psi= \d_\pm \hat \psi \pm
d_\pm [E_\pm,\hat \psi]$. $\hat \psi$ can be chosen in $H_{taf}$. 
Conjugating the original spectral parameter $t$ by $\psi_V$ allows to 
reintroduce a ``moving spectral parameter" $m$, moving with respect to $t$.
In \cite{julanico} moving and fixed have been exchanged from the original
\cite{mai1} point of view to the present one.  
$m:=\psi_V\cdot t\cdot \psi_V^{-1}$.
This notation simply encodes the action of the Virasoro reparameterisation
factor on the Kac-Moody factor of the semidirect product. This will bring us 
into the Geroch group formalism of \cite{mais} in section 4.4.

By  factorizing the wave function in the reverse order,
the linear system can be projected into the affine algebra in another way.  
Namely, write $\Psi=\psi_V\cdot\psi$, then $\psi$
is solution of the following linear system:
\debut
(\d_\pm+  A^{km}_\pm)~\psi=0\quad
{\rm with} \quad  A^{km}_\pm= \psi_V^{-1}~{\hat A^{km}_\pm}~\psi_V 
 \label{Akm}
\fin
The connection of that linear problem belongs to the affine Kac-Moody algebra
 $\CH_{taf}$ of $t':=\psi_V^{-1}\cdot t\cdot \psi_V$.

\subsection{Chiral fields}

The existence of chiral fields follows from the fact that the 
components $A_+$ and $A_-$ belong to the two Borel subalgebras
$(\CB_\pm \oplus \Cmath k \oplus \Cmath E_\pm) $ and from the existence 
of highest weight vectors \cite{babono}. 

Let $\La$ be a highest weight of $V \CH$, and
$\ket{\la}$ be vectors of the highest representation of $\La$.
Let us parametrize $Q_\pm$ as $Q_\pm = U_\pm \d_\pm U^{-1}_\pm$ with $U_\pm\in H$.
Only the product $U^{-1}_-U_+$ is gauge invariant, since
under a gauge transformation (\ref{gauge}), $U_\pm\to\La U_\pm$.
The fields $\bar \xi_{\la}$ and $\xi_{\la}$ defined below are chiral:
\debut
\bar \xi_{\la}(z)= \rho^{-h_\La}e^{\hat \sig k/2}~\cdot
\bra{\la}~\({U^{-1}_-\cdot \Psi}\) \quad &;&\quad \d_-\bar \xi_{\la} =0 
\label{xibar}\\
\xi_{\la}(z) = \({\Psi^{-1}~ U_+}\)~\ket{\la}~\cdot 
\rho^{-h_\La}e^{\hat \sig k/2}
\quad &;&\quad \d_+ \xi_{\la} = 0 
\label{xi}
\fin
Since under gauge transformations $U_\pm \to \La U_\pm$ and 
$\Psi \to \La\Psi$, the chiral fields $\bar \xi_{\la}$ and $\xi_{\la}$ are
invariant under gauge transformations.
Moreover, if $\ket{\la_1}$ and $\ket{\la_2}$ are two
vectors of the highest representation of $\La$, the scalar product of
$\bar \xi_{\la_1}$ and $\xi_{\la_2}$ gives:
\debut
\bar \xi_{\la_1}\cdot \xi_{\la_2}
= \rho^{-2h_\La}~ e^{\hat \sig k}~\cdot 
\bra{\la_1}(U_-^{-1}U_+) \ket{\la_2} \label{xixi}
\fin
We can thus reconstruct the conformal factor $\hat \sig$ and
the  gauge invariant part of $Q_\pm$ from the chiral fields.
The scalar product (\ref{xixi}) is directly related to the $\tau$-functions
$\tau_\pm^0$ we will introduce in Section 5. If we insert the 
specific operator $\hat \CE^*$ between the chiral fields before taking
the scalar product, ie. considering $\bar \xi_{\la_1}\cdot \hat \CE^*\cdot \xi_{\la_2}$,
we could reproduce quantities appearing in (\ref{003}).

Let us sketch the proof of eq.(\ref{xibar}).
Differentiating $\bar \xi_\la$ using the fact that $\Psi$ is
solution of the linear system, we get:
\debut
\d_-\bar \xi_\la= -h_\La d_- \bar \xi_\la +
(\d_-\hat \sig)\bar \xi_\la \frac{k}{2} +
\rho^{-h_\La}e^{\hat \sig k/2}~\bra{\la}U^{-1}_-(Q_--A_-)\Psi \non
\fin
Now, $\bra{\la}U^{-1}_-$ is a vector of the highest representation of $\La$,
therefore $$\bra{\la}U^{-1}_-(Q_--A_-)=
\bra{\la}U^{-1}_-(h_\La d_- - (\d_-\hat \sig)\frac{k}{2}).$$
This proves that $\d_-\bar \xi_\la=0$. The equation
$\d_+ \xi_{\la} = 0$ is proved in a similar way.

\def\bu{ {\bf u}}

\section{Geroch and dressing symmetries.}
In this section we define and describe the dressing transformations.
We then compare them to the usual Geroch transformations.

\subsection{Dressing symmetries.}
We restrict ourselves to dressing by elements in $H_{taf}$. The extension 
to dressing by elements which are exponentials
of $E_\pm$ is simple; it is discussed in App. C in the case where the 
wave function belongs to the two dimensional extension of $H_{taf}$ by $E_\pm$.
But in this section we choose the normalization condition such that the wave 
function may belong to $\bH$.
The dressing transformations are non-local gauge transformations
which preserve the form of the Lax connection \cite{zakasha,zaka,semen}, see 
also
\cite{ueno,babe}. For any given element $g= g_-^{-1}g_+\in H_{taf}$, factorised 
according to \ref {factor2}, they are 
defined by:
\debut
\Psi \to D_{(g_-,g_+)}(\Psi)=(\Psi g\Psi^{-1})_-\cdot\Psi \cdot g_-^{-1} 
= (\Psi g\Psi^{-1})_+\cdot\Psi \cdot g_+^{-1}
\label{dress}
\fin
with $(\Psi g\Psi^{-1})_-^{-1} (\Psi g\Psi^{-1})_+=\Psi g\Psi^{-1}$. 
On the Lax connection they act as non-local gauge transformations:
\debut
D_{(g_-,g_+)}(A)= \Th_\pm A \Th_\pm^{-1} - \d\Th_\pm\cdot \Th_\pm^{-1}
\label{dreslax}
\fin 
with $\Th_\pm=(\Psi g\Psi^{-1})_\pm \in H_{taf}$, ie.
$\Th_-^{-1}\Th_+=(\Psi g\Psi^{-1})$, because $(\Psi g\Psi^{-1})$ is still in
$H_{taf}$.
These non-local gauge transformations should not be confused with
the gauge transformations associated to the finite dimensional 
group $H$. They act on the physical fields.
In eqs.(\ref{dress},\ref{dreslax}) we can choose to implement the transformation
either with $\Th_+$ or with $\Th_-$, the results are identical.
For $\hat X=\hat X_+-\hat X_-\in \CH_{taf}$, the infinitesimal
dressing transformations are:
\debut
D_{(\hat X_-,\hat X_+)}(\Psi)&=& \hat Y_\pm \Psi -\Psi \hat X_\pm ,
\quad {\rm with}\quad
\hat Y_\pm= (\Psi\hat X\Psi^{-1})_\pm\in (\CB_\pm \oplus \Cmath k) \non \\
D_{(\hat X_-,\hat X_+)}(A)&=& -\d \hat Y_\pm -[A,\hat Y_\pm] 
\label{dresinf}
\fin

These transformations are gauge covariant. Indeed, under a gauge transformation,
$\Psi\to \Psi^\La=\La\Psi$, thus $(\Psi g\Psi^{-1})\to \La (\Psi g\Psi^{-1})\La^{-1}$,
and $\Th_\pm\to \La \Th_\pm\La^{-1}$. This implies that
$ D_{(g_-,g_+)}(\Psi^\La)= \La  D_{(g_-,g_+)}(\Psi)$. 
\bigskip

The two important points concerning the transformations (\ref{dreslax}) is
that they preserve the zero curvature condition since they are
gauge transformations and secondly that they preserve the form of the Lax 
connection, as we will soon show.  Therefore, they map solutions of the 
equations of motion into new solutions.
\bigskip

Let us prove that the form of the Lax connection is preserved by the
transformation (\ref{dreslax}). As usual, the trick is that the
transformed Lax connexion $D_{(g_-,g_+)}(A)$ can be obtained 
in two different ways: using either $\Th_+$ or
$\Th_-$. Consider for example $D_{(g_-,g_+)}(A_+)$.
First since $\Th_\pm\in H_{taf}$ the projection of $D_{(g_-,g_+)}(A_+)$
on the Virasoro algebra is equal to that of $A_+$. Ie. $\rho$
is untouched by the dressing. Next we have to prove that the projection
of $D_{(g_-,g_+)}(A_+)$ on $\CH_{taf}$ decomposes on elements of 
degree zero or one, ie. on elements of the form $(X\otimes t^0)$
and $(X\otimes t^1)$, and on the central element $k$. First let us
implement the gauge transformation using $\Th_+$. Since $\Th_+\in
\exp(\CB_+\oplus \Cmath k)$, this gauge transformation increases the degree
and therefore $D_{(g_-,g_+)}(A_+)$ decomposes on elements of positive
degree. Similarly, implementing the gauge transformation using
$\Th_-\in \exp(\CB_-\oplus \Cmath k)$ which decreases the degree shows that
$D_{(g_-,g_+)}(A_+)$ decomposes on elements of degree less that one.
Note that $\Th_- E_+\Th_-^{-1}-E_+$ which naively is of degree less than two
is actually of degree less than one, as it should be.
Altogether, it implies that $D_{(g_-,g_+)}(A_+)$ decomposes on the central
element and on elements of degree zero or one. Similarly, one
shows that $D_{(g_-,g_+)}(A_-)$ decomposes on elements of
degree zero or minus one and on the central element.
Lastly, one has to verify that there exists $\hat \sig'$ such that
the projections on the central element of $D_{(g_-,g_+)}(A_\pm)$ 
are $\mp(\d_\pm \hat \sig')$. Doing the gauge transformation with
$\Th_+$ for $A_+$ and with $\Th_-$ for $A_-$ one checks that this
condition is satisfied provided we specify, as we did, the factorization 
problem such that $\Th_\pm$ have opposite components on the central element.

\bigskip

Moreover, this proof gives the explicit expression for the dressed
fields. Namely,  for $g=g_-^{-1}g_+\in H_{taf}$, let us decompose
$\Th_\pm$ as:
\debut
\Th_\pm := (\Psi g\Psi^{-1})_\pm 
= e^{\pm \zeta k/2}\cdot  h_\pm(0)h_\pm(1)\cdot M_\pm , 
\label{hzero}
\fin
with
\debut
h_\pm(0)\in H,~~~~~~~~ h_\pm(1)=\exp(X_{(\pm1)}\otimes t^{\pm1}) 
\label{hzero2}
\fin
where $X_{(\pm1)}\in {\bf r}$
and $M_+$ ($M_-$) with degree bigger than two (less than minus two), 
ie. $M_\pm$ is the exponential of elements of the form $(X\otimes t^{\pm n})$
with $n\geq 2$. Then, by doing the transformation (\ref{dreslax})
 either with $\Th_-$ or $\Th_+$ as explained above, one finds that
the fields $Q_\pm$,  $P_\pm$, $\hat \sig$ and $\rho$ are transformed according to:
\debut
D_{(g_-,g_+)}(Q_\pm) &=& h_\pm(0)Q_\pm h_\pm(0)^{-1}
-\d_\pm h_\pm(0)\cdot h_\pm(0)^{-1} \label{drq}\\
D_{(g_-,g_+)}(P_\pm) &=& h_\mp(0)\({ P_\pm+\frac{d_\pm}{2}X_{(\mp1)} }\) h_\mp(0)^{-1} 
\label{drp}\\
D_{(g_-,g_+)}(\hat \sig) &=& \hat \sig + \zeta \label{drsig}\\
D_{(g_-,g_+)}(\rho) &=& \rho \label{drrho}
\fin
The formula (\ref{drp}) for $D_{(g_-,g_+)}(P_\pm)$ have been
obtained by implementing the dressing transformation with $\Th_\mp$.
Choosing on the  contrary to implement it with $\Th_\pm$ gives
alternative but equivalent expressions for $D_{(g_-,g_+)}(P_\pm)$:
\debut
D_{(g_-,g_+)}(P_\pm)= h_\pm(0)\({ P_\pm +[X_{(\pm1)},Q_\pm]
+\frac{d_\pm}{2}X_{(\pm1)}- \d_\pm X_{(\pm 1)} }\) h_\pm(0)^{-1}
\label{drpbis}
\fin
Thus, dressing transformations act on $Q_\pm$
as non-local chiral gauge transformations.
As it should be,  only the product $h(0)=h_-(0)^{-1}h_+(0)$
matters up to gauge transformations.
For example we can choose that  $Q_-$ remains unchanged by applying a 
compensating local gauge transformation, we obtain then:
\debut
D_{(g_-,g_+)}(Q_-)&=& Q_- \label{dresq1}\\
D_{(g_-,g_+)}(Q_+) &=&  h(0) Q_+ h(0)^{-1} - \d_+ h(0)\cdot h(0)^{-1}
\label{dresq} \\
D_{(g_-,g_+)}(P_-) &=& h(0)\({ P_-+\frac{d_-}{2}X_{(+1)}}\) h(0)^{-1} \label{dresp1}\\
D_{(g_-,g_+)}(P_+)&=&P_+ +\frac{d_+}{2}X_{(-1)} 
\label{dresp}
\fin
We recall that, if $Q_\pm$,  $P_\pm$, $\hat \sig$ are
solutions of the equations of motion, then
$D_{(g_-,g_+)}(Q_\pm)$, $D_{(g_-,g_+)}(P_\pm)$ and $D_{(g_-,g_+)}(\hat \sig)$
are new solutions.

We leave as an exercise to the reader to
verify that the dressing of the chiral fields is very simple.
Namely,
\debut
D_{(g_-,g_+)}(\bar \xi_\la)= \bar \xi_\la\cdot g_-^{-1}
\quad;\quad
D_{(g_-,g_+)}(\xi_\la)= g_+\cdot \xi_\la \label{dresxi}
\fin
Ie. the $g_\pm$ components of $g$ act separately on the two chiral sectors.
This is very analogous to what happens in the Toda field theories \cite{babe}.
\bigskip
A few examples of dressing transformations are given in Appendix B.
 
Finally, let us write explicitly the dressing transformations in the
case $H\bacs G=SO(2)\bacs SL(2,\Rmath)$. Let us parametrize the connection
$(Q_\pm+P_\pm)$ as:
$Q_\pm =  q_\pm (\sig^+-\sig^-)$,
$P_\pm =  p^1_\pm\sig^z +  p^2_\pm (\sig^++\sig^-)$.
We chose  $Q_-$ to  remain unchanged.
Let us parametrize $h(0)$ by
$h(0)=\exp(-\frac{\varphi}{2}(\sig^+-\sig^-))$. Then:
\debut
D(q_-) = q_- \quad&;&\quad D(q_+)=q_+ +\half\d_+\varphi \label{dress2.1}\\
D\pmatrix{p_-^1 \cr p^2_- \cr} =
\pmatrix{\cos\varphi & -\sin\varphi\cr \sin\varphi &\cos\varphi\cr}
\pmatrix{p_-^1+\frac{d_-}{2}X_{(+1)}^1\cr p^2_-+\frac{d_-}{2}X_{(+1)}^2\cr} \quad&;&\quad
D\pmatrix{p_+^1\cr p^2_+\cr} =
\pmatrix{p_+^1+\frac{d_+}{2}X_{(-1)}^1\cr p^2_++\frac{d_+}{2}X_{(-1)}^2\cr} \label{dressn.2}
\fin
where we have decomposed $X_{(\pm1)}$ as $X_{(\pm1)}=
X^1_{(\pm1)}\sig^z + X^2_{(\pm1)} (\sig^++\sig^-)$.
Recall that $\hat \sig$ is dressed according to (\ref{drsig}).
\bigskip
 
\subsection{Right action of $\bf VH$ and its commutation relations with the dressing group.}
We can also define an action of $\bf VH$
on the wave functions by acting on the right by constant elements of $\bf VH$:
\debut
\Psi \to R_g(\Psi)=\Psi \cdot g^{-1}, \quad  g \in {\bf VH} \label{right}
\fin
These transformations leave the Lax connection invariant,
ie. $Q_\pm$, $P_\pm$ and $(\d_\pm\hat \sig)$
are unchanged, and the $n$-bein $E$ is unmodified
up to global $H$ transformations.
The chiral fields transform as $R_g(\bar \xi_\la)=\bar \xi_\la\cdot g^{-1}$
and $R_g(\xi_\la)=g\cdot \xi_\la$.
Note that we defined the right action for any element of the semi-direct
product of the Virasoro group by $H_{taf}$ and not only
for elements of $H_{taf}$ as we did for dressing transformations.
\bigskip

The right actions and the dressing transformations should not be confused.
They do not commute.  Their commutation relations are:
\debut
D_{(g_-,g_+)} D_{(h_-.h_+)}&=& D_{(g_-h_-,g_+h_+)} \non\\
R_g R_h &=& R_{gh} \label{comm1}\\
R_h^{-1}D_{(g_-,g_+)} R_h &=& R_{(h^{-1}g_\pm h)(h^{-1}gh)_\pm^{-1}}
\cdot D_{((hgh^{-1})_-,(hgh^{-1})_+)}\non
\fin
As we explained, the factorization problem has been specified up to
the gauge freedom $g_\pm \to h g_\pm$, $h\in H$. The commutation
relations are written here for a given and fixed choice of this
gauge freedom.

The Lie brackets of the infinitesimal transformations are: 
\debut 
\[D(\hat X_-,\hat X_+),D(\hat Y_-,\hat Y_+)\] &=& 
D([\hat X_-,\hat Y_-],[\hat X_+,\hat Y_+]) \non\\  
\[R(\hat X),R(\hat Y)\] &=& 
R([\hat X,\hat Y]) \label{comm2}\\ 
\[D(\hat X_-,\hat X_+),R(\hat Y)\] &=& 
D([\hat X,\hat Y]_-,[\hat X,\hat Y]_+) 
+ R([\hat X_\pm,\hat Y]-[\hat X,\hat Y]_\pm) \non 
\fin
where $\hat X$ is decomposed according to the factorization problem
as $\hat X= \hat X_+- \hat X_-$ with $\hat X_\pm \in (\CB_\pm \oplus Ck)$.
We recall again that dressing transformations are defined for
$\hat X= \hat X_+- \hat X_-\in \CH_{taf}$ whereas right actions
are defined for any element in the semi-direct product
$V\CH= Vir \times \CH_{taf}$. In the last commutator the dressing 
term on the right is in $\CH_{taf}$.

The above commutation relations of the dressing transformations are well known,
cf eg. \cite{semen,babe}. They imply that the $g_-$ and $g_+$ components in
the factorized elements $g=g_-^{-1}g_+$ act separately.
In other words, in the dressing algebra the two Borel
subalgebras $(\CB_\pm \oplus \Cmath k)$ commute.
The commutation relations between the right actions and the
dressing transformations directly follow from their definitions.
They imply that the full algebra is generated by right multiplications
$R(\hat X),~ \hat X\in V\CH$ and dressing transformations by 
elements in $\bf h$, ie $D(X\otimes t^0)$.
 
\bigskip
{\bf To summarize}: dressing transformations and right actions form
an infinite dimensional solution generating group for Einstein's equations
with commutation relations (\ref{comm1}).

\bigskip

\subsection{Geroch group as dressing transformations on the connection.}
Let us consider the gravity case, $H\bacs G=SO(2)\bacs SL(2,\Rmath)$.
The Geroch group arises by combining the $SL(2)$ global group
with a duality transformation.
In order to compare it with dressing transformations,
we are first going to describe its action on the Lax connection.
\bigskip

Let us first recall a few basic facts concerning the Geroch group \cite{ger},
cf. eg. \cite{nico, mais, julia}.
Recall the equations of motion (\ref{Etri1},\ref{Etri2}) in
the triangular gauge. The first equation (\ref{Etri1}) is
a conservation law for the current $\rho\De^2\d_\pm N$.
This current can be dualized by introducing another field
$N^*$ such that $\rho\De^2\d_\pm N = \pm \d_\pm N^*$.
More precisely, let us introduce the dual variables
$N^*$,  $\De^*$ and $\hat \sig^*$ by:
\debut
\De^* &=& \inv{\rho \De} \label{dual1}\\
\De^*~ \d_\pm N^* &=& \pm \De~ \d_\pm N \label{dual2}\\
\De^*~ e^{4\hat \sig^*} &=& \De~ e^{4\hat \sig} \label{dual3}
\fin
It defines $N^*$ up to an additive constant.
The dual metric is:
\debut
ds^2_* &=& 2 \rho^{\half} e^{2\hat \sig^*} da db +
\rho \De^{*~-1} dx^2 + \rho \De^*(dy - N^* dx)^2 \non\\
 &=& 2 \rho\De e^{2\hat \sig} da db + 
\rho^2 \De dx^2 + \De^{-1}(dy - N^* dx)^2 
\label{metric4}
\fin
We will denote by $\al$ this transformation:
$\al(\De, N, \hat \sig) =(\De^*, N^*, \hat \sig^*)$ .
It is well-defined only once the integration constant
in eq.(\ref{dual2}) has been fixed.
For the connection, it corresponds to the simple transformation:
\debut
Q_\pm+P_\pm &=& \half(\De^{-1}\d_\pm\De)\cdot\sig^z + \De(\d_\pm N)\cdot\sig^-\non\\
\to\qquad 
Q^*_\pm+P^*_\pm &=& -\half(\rho^{-1}\d_\pm\rho)\cdot\sig^z
-\half(\De^{-1}\d_\pm\De)\cdot\sig^z \pm \De(\d_\pm N)\cdot\sig^-\non 
\fin
Notice that the duality transformation $\al$ is well defined 
on the connection $(Q_\pm+P_\pm)$ independently of the
integration constant needed to define $N^*$ with eq.(\ref{dual2}).
\bigskip

The important point is that
if the variables $N,~\De$ and $\hat \sig$ are solutions
of the equations (\ref{Etri1},\ref{Etri2}), then
the dual variables $ (\De^*, N^*, \hat \sig^*)$
also form a solution of these equations.
In other words, the theory is auto dual.
\bigskip

By definition, the Geroch group is the group generated by the two
groups of global $SL(2,\Rmath)$ transformations on the two sets
of dual variables. It requires the choice and definition of an infinite set of 
dual potentials. 

The global $SL(2,\Rmath)$ transformations
in the original variables $\De, N, \hat \sig$ is the 
``Matzner-Misner" group with generators $J_z,~ J_\pm$
that we have considered in the previous section.
It was defined by acting on the right on the zwei-bein.
Since it leaves  $\hat \sig$ and the connection $(Q_\pm+P_\pm)$ invariant
up to gauge transformations, it also leaves the Lax connection $A_\pm$ 
invariant up to gauge transformation. 
\bigskip

The $SL(2,\Rmath)$ transformations
in the dual variables $\De^*, N^*, \hat \sig^*$ is
called the ``Ehlers" group. We denote its generators by
$J_z^*,~ J_\pm^*$. If we would have fixed the normalization
constant entering in the definition (\ref{dual2}), we then 
would have defined these transformations using the 
the duality transformation $\al$, ie.:
\debut
J_z^*= \al^{-1}~J_z~\al,\quad J_\pm^*=\al^{-1}~J_\pm~ \al \label{dualJ}
\fin
So it is simpler to first define the action of the
Ehlers transformation on the Lax connection, ie. on $(Q_\pm +P_\pm)$
and $(\d_\pm \hat \sig)$, and then to integrate them
to define  the action on $(\De, N,\hat \sig)$.
The action on the Lax connection in the triangular gauge is :
\debut
J_z^*(Q_\pm+P_\pm)= 0\quad &;&\quad J_z^*(\d_\pm \hat \sig)= 0  \non\\
J_-^*(Q_\pm+P_\pm)= 0 \quad &;&\quad J_-^*(\d_\pm \hat \sig)= 0\non\\
J_+^*(Q_\pm+P_\pm)= \mp\rho\De^2(\d_\pm N)\cdot \sig^z \pm
2\d_\pm( \rho\De)\cdot \sig^- \quad &;&\quad 
J_+^*(\d_\pm \hat \sig) = \pm \rho\De^2(\d_\pm N)\label{jp*}
\fin
$J_z^*$ and $J_-^*$ act trivially on the Lax connection,
only $J_+^*$ does not act trivially.
One may directly check that it induces a symmetry of
the equations of motion.
On the Lax connection in the triangular gauge these actions are
local also in terms of the original variables $(\De, N, \hat \sig)$.
\bigskip

Let us now show that these transformations are special dressing
transformations. Thus we have to prove that the infinitesimal
``Ehlers'' transformation (\ref{jp*}) can be presented as a
non-local gauge transformation acting on the Lax connection, ie:
\debut
J^*_+(A_\pm) = -\d_\pm\hat Y - [A_\pm,\hat  Y]
\label{ehlersinf}
\fin
with $z_\pm$-dependent elements $\hat Y \equiv \hat Y^*_\pm$ which respectively belong to 
$(\CB_+\oplus \Cmath k)$ or to $(\CB_-\oplus \Cmath k)$. In the triangular gauge
(\ref{QPtri}), the variation $J^*_+(A_\pm)$
of the Lax connection corresponding
to the transformation (\ref{jp*}) is:
\debut
J^*_+(A_\pm) &=& -\rho\De^2(\d_\pm N)\cdot\frac{k}{2}
 \mp \d_\pm(\rho\De)\cdot(\sig^+-\sig^-)  \non\\
&~&\pm \d_\pm( \rho\De)\cdot (\sig^++\sig^-)\otimes t^{\pm1}
\mp \rho\De^2(\d_\pm N)\cdot \sig^z\otimes t^{\pm1} 
\label{j*a}
\fin
The linear system (\ref{ehlersinf}) is compatible, since the
statement that $J^*_+$ generates a symmetry of the equations
of motion is equivalent to the compatibility condition
$D_-^A(J^*_+(A_+))-D^A_+(J^*_+(A_-))=0$ with $D^A_\pm=\d_\pm+A_\pm$.
However, to prove that $J^*_+(A_\pm)$ is a dressing 
transformation, we have to look for a solution $\hat Y^*$ of (\ref{ehlersinf})
which either belongs  to $(\CB_+\oplus \Cmath k)$ or to $(\CB_-\oplus \Cmath k)$.
For the special form (\ref{j*a}) of the variation $J^*_+(A)$ there exist
two solutions $\hat Y^*_\pm$ respectively in $(\CB_\pm \oplus \Cmath k)$. 
Their explicit expressions in the triangular gauge are:
\debut
\hat Y^*_\pm = \pm N^* \frac{k}{2}
\pm (\rho \De)\cdot (\sig^+-\sig^-)\otimes\({\frac{1+t^{\pm2}}{1-t^{\pm2}}}\)
\mp (\rho\De)\cdot (\sig^++\sig^-)\otimes\({\frac{2t^{\pm1}}{1-t^{\pm2}}}\)
\label{y*pm}
\fin
Note that the components of $\hat Y^*_\pm$
on the central element are opposite as it should be.
We could have expected the existence of this dressing transformation
on the ground that the phase space contains only one dressing orbit.

The last step in proving that $J^*_+(A)$ is a dressing transformation
consists in proving that the two solutions $\hat Y^*_\pm$ 
can be written as $\hat Y^*_\pm = (\Psi \hat X^* \Psi^{-1})_\pm$ for
some $z_\pm$-independent $\hat X^* \in \CH_{taf}$. Since both
$\hat Y^*_+$ and $\hat Y^*_-$ are solutions of the same linear system,
eq.(\ref{ehlersinf}), their difference $\hat Y^{*}=\hat Y^*_+
-\hat Y^*_-$ is solution of \debut
\d_\pm \hat Y^{*} + [A_\pm, \hat Y^{*}]=0 \non
\fin
Hence, there formally exists a constant element $\hat X^* \in \CH_{taf}$,
such that $\hat Y^{*}= \Psi \hat X^* \Psi^{-1}$. Since by construction
$\hat Y^*_\pm$ are the $\pm$~~components of $\hat Y^{*}$ as defined
by the factorization problem, this implies that 
$\hat Y^*_\pm = (\Psi \hat X^* \Psi^{-1})_\pm$.
One should handle $\hat Y^*$ skillfully since it is a sum of two
formal power series with inverse argument.
It will be identified with a vertex operator in Section 6.3.
In particular the conjugation by $\Psi$ may make sense only
inside appropriate matrix elements.
\bigskip

It is surprising that $\hat Y^*_\pm$ can be expressed simply in terms of
the dual variables $N^*$ and $\De^*=1/(\rho\De)$.
It means that the dual variables are probably calling for a more
algebraic interpretation. This will be used in Section 5.3 to present an
algebraic way of evaluating the dual fields.
It is also intriguing to note that $\hat Y^*_\pm$ are very closely related to
the elements which by dressing generate solitons in the sine-Gordon
theory \cite{babe2}. 
\bigskip

\subsection{Lifting the Geroch transformations to the wave function.}
In the previous section we essentially described the echo
of Geroch transformations on the Lax connection. We now 
want to describe their action on $(\De, N, \hat \sig)$.
For this we need to lift them to the wave function.
This discussion will be close to the usual presentation
of Geroch transformations \cite{mais}, but it will allow us
to compare  more precisely Geroch transformations 
with dressing transformations.
This will force us to consider wave functions outside $\bf {VH}$.
It will be convenient to change the parameter by using $s$ 
instead of $t$ with:
\debut
s= \frac{1-t}{1+t}, \quad t=\frac{1-s}{1+s} \label{defs}
\fin
Note that this map exchanges halfplane and unit disk of the parameters.
In the $s$-parametrisation the vector fields
$E_\pm$ are $E_\pm =\pm (\frac{1\mp s}{1\pm s}) s\d_s$.
We define $\CG^s_{af}$ to be the affine algebra with spectral parameter
$s$ generated by elements of the form $X\otimes s^n$ with $X\in \bf g$ 
and $n$ integer with commutation relations:
\debut
\[{ X\otimes s^n, Y\otimes s^m }\]=
[X,Y]\otimes s^{n+m} +n \frac{k'}{2} ~tr(XY)\de_{n+m,0} \label{comhs}
\fin
The central charge $k'$ is not the same as in the $t$ parametrization. 
Let $G^s_{af}$ the corresponding group.
In this parametrization the automorphism $\tau'$  is now:
\debut
\tau'(X\otimes s^n) = \eta(X)\otimes s^{-n}
\quad , \quad \tau'(k') = - k'
\label{invo2}
\fin
Notice that $\tau'$ coincides with $\tau$ on the generators of the
loop subalgebra but not on the central charges.
We denote by $\CK^s_{af}$ the $\tau'$ invariant subalgebra of $\CG^s_{af}$
and by $K^s_{af}$ the corresponding group.
Remark that $k'$ does not belong to $\CK^s_{af}$.
Loosely speaking $\CK^s_{af}$ is the twisted loop algebra $\CH_{taf}$
but reexpressed with the spectral parameter $s$.
The bilinear form defined by $(X\otimes s^n,Y\otimes s^m)'=
tr(XY)\de_{n+m,0}$ is negative definite on $\CK^s_{af}$.
So, $K^s_{af}$ may be interpreted as the maximal  compact subgroup of 
$G^s_{af}$ \cite{julia}.

For the sake of simplicity we will deal mostly with the loop algebra
and therefore forget about the central charges. 
In particular, the Lax connection will belong to the semi-direct
product of $Vir$ by the loop algebra, and the wave function
$\hat \psi_G \equiv \Psi_G\Psi_0^{-1}$ will belong to the loop group. 
However proving an isomorphism between the loop algebras  with 
different spectral parameters $s$ and $t$ would require
a more mathematical discussion than the following.
As a consequence, some of the manipulations we are going to use
may require more precise definitions and appropriate completions.

The usual formulation of the Geroch transformations starts
by imposing a normalisation  condition, or a ``triangular gauge", 
on the wave function $\Psi_G$. In our notation, it reads:
\debut
\hat \psi_G \equiv \Psi_G\cdot \Psi_0^{-1} = {\rm regular~at~} (s=0)
\label{condi}
\fin
This forces  $\hat \psi_G$ to belong to the full (untwisted) affine
group $G^s_{af}$ including $k'$, and thus $\Phi_G$ belongs to semi direct product
of $Vir$ with $G^s_{af}$. 
This choice which is  unnatural (from the dressing point 
of view) will be made only in this Section. 
Following \cite{mais}, we assume that the condition
(\ref{condi}) can be consistently imposed; it requires
that $\hat \psi_G$ has a non trivial component along the
central generator $k'$.
This regularity condition at $s=0$, ie at $t=1$, ensures 
that the zweibein $\CV$ can be reconstructed by expanding 
$\hat \psi_G$ in powers of $s$:
\debut
\hat \psi_G = \CV + \CO(s) ~~\in G^s_{af}\non
\fin
We also impose the triangular gauge on $\CV$ as in eq.(\ref{Etri}).
Note that consistency of this gauge condition requires
to expand all functions of $s$ entering the Lax connection
in positive powers of $s$. In particular, the vector fields $E_\pm$
have to be considered as regular at $s=0$.

Now there is an action of the untwisted affine algebra $\CG^s_{af}$
with spectral parameter $s$ on the wave functions which is
analogous to the action of $G$ on the coset space $K \bacs G$. One must now
use the Iwasawa decomposition to parametrise the set of configurations by 
triangular elements like $\hat \psi_G$.  
The infinitesimal generators of $\CG^s_{af}$ can be described as follows.
First we act on $\Psi_G$ on the right by the constant elements in $\CG^s_{af}$
of the form $\hat X_s=X\otimes s^n$, with $n\in \Zmath$, $X\in \bf g$.
This may violate the triangular gauge condition (\ref{condi})
so we restore it by acting on the left with elements $\hat Y(\Psi_G,\hat X_s)$
in $\CK^s_{af}$. Namely:
\debut
\delta\Psi_G = \hat Y(\Psi_G,\hat X_s)\cdot \Psi_G - \Psi_G\cdot \hat X_s
\label{affGH}
\fin
The transformation of the Lax connection induced by eq.(\ref{affGH})
is simply a non-local gauge transformation with parameter $\hat Y(\Psi_G,\hat X_s)$.
The induced action on $\hat \psi_G$ is:
\debut 
\delta\hat \psi_G = \hat Y(\Psi_G,\hat X_s)\cdot \hat \psi_G - 
\hat \psi_G\cdot \hat X_v 
\label{affGH2} 
\fin
where $\hat X_v= \Psi_0\cdot \hat X_s\cdot \Psi_0^{-1}$. 
They are of the form $\hat X_v= X\otimes v^n$ with
$v= \Psi_0\cdot s\cdot \Psi_0^{-1} = (\frac{s}{\rho})[1+\CO(s)]$.
This defines $v$ as a formal power series in $s$.
As meromorphic function of $s$, $v$ has square root branch cut
because the group $Tr^2$ acts homographically on the variable $q=\frac{2s}{s^2+1}$.
It satisfies $\d_\pm v+ d_\pm (\frac{1\mp s}{1 \pm s}) s\d_s v =0$.
Let us connect these definitions with the previous ones as reviewed in
\cite{julanico}. The relevant parameter there was called $y$:   
$2/y:=a(1/s+s-2)+b(1/s+s+2)$, and $v$ is nothing but $y$. 

In other words, the action on $\hat \psi_G$ is constructed by acting on the 
right by elements of $G_{af}$ but with the ``moving" spectral parameter $v$.
The compensating factor $\hat Y(\Psi_G,\hat X_s)$, which 
has to be choosen in $\CK^s_{af}$, is determined
by requiring that the singularities of $\delta\hat \psi_G$
at $s=0$ vanish. Thus
\debut
\hat Y(\Psi_G,\hat X_s) =  (1 + \tau')~\inv{2i\pi}\oint_{|s'|<|s|} 
\frac{ds'}{s-s'} (\hat \psi_G\cdot \hat X_v\cdot \hat \psi_G^{-1})(s') 
\label{YH}
\fin

It is simple to find the action of the simple root generators of $\CG^s_{af}$.
The generators of the form $X\otimes s^0$ do not spoil the regularity
condition at $s=0$. A compensating factor is needed only when
the triangular gauge condition imposed on the zweibein is threatened. 
So they act in the same way as they were acting on the zweibein.
The generator $\sig^+\otimes s$ preserves the triangular gauge as well
and therefore does not require any compensating factor.
The central charge acts by translation of the conformal factor.
Finally the generator $\sig^-\otimes s^{-1}$ breaks the regularity
condition and therefore requires a compensating 
factor $\hat Y(\Psi_G,\sig^-\otimes s^{-1})$.
It can be computed explicitely using eq.(\ref{YH}) and the 
expression (\ref{Etri}) of the zweibein:
\debut
\hat Y(\Psi_G,\sig^-\otimes s^{-1})
= -(\rho \Delta)~[ \sig^+\otimes s - \sig^-\otimes s^{-1} ] 
\label{YH2}
\fin
One may check that the transformation of the Lax connection
induced by $\hat Y(\Psi_G,\sig^-\otimes s^{-1})$ is the same as  the 
Ehlers transformation (\ref{jp*}) with $k$ set to zero.

Thus the generators of the form $X\otimes s^0$ act as the generators
of the Matzner-Misner group, while the generators
$k'-\sig^z\otimes s^0$, $\sig^+\otimes s$ and $\sig^-\otimes s^{-1}$
act as the generators of the Ehlers group.
This shows that the Matzner-Misner and Ehlers groups generate
the untwisted affine $SL(2,\Rmath)$ Kac-Moody $G^s_{af}$ \cite{julia, mais}. 
Eq.(\ref{affGH}) describes the infinitesimal Geroch transformations,
the non infinitesimal ones are formally written as:
\debut
\Psi_G \to K^s(\Psi_G,e_s)\cdot \Psi_G\cdot e^{-1}_s \label{glob}
\fin
with $e_s$ constant elements of $G^s_{af}$ and $K^s=K^s(\Psi_G,e_s)$
the compensating factors in $K^s_{af}$.

How can we make contact with dressing transformations?
Since the central charge $k$ is not an element of $G^s_{af}$
we will project it out.
The point is that given the compensating factor
$\hat  Y(\Psi_G,\hat X_s)$ as a function of $s=\frac{1-t}{1+t}$,
it can be expanded either in positive or in negative powers of $t$.
Calling $\hat Y_\pm$ these two expressions of $\hat  Y(\Psi_G,\hat X_s)$.
The gauge transformations associated to the Geroch transformations
can therefore be implemented in the $t$-parametrisation 
in two ways, either with $\hat Y_+$
or with $\hat Y_-$. These gauge transformations are those used
in the dressing transformations. This explains why the 
transformations (\ref{affGH}) preserve the form of the Lax
connection and therefore are symmetries of the equations
of motion.  As a check one may verify that
the elements $\hat Y^*_\pm$ determined in eq.(\ref{y*pm}) with $k=0$
 are indeed different $t$-expansions of the function 
$\hat Y(\Psi_G,\sig^-\otimes s^{-1})$.
This construction thus explains the simple form of $\hat Y^*_\pm$ (with $k=0$)
found in eq.(\ref{y*pm}) which seemed at first surprising.

In other words, given an element $Y^s$ of the ``compact" subalgebra $\CK^s_{af}$,
which is by definition a polynomial function of $s$ and $s^{-1}$,
we associate, by expanding it either in positive or negative
powers of $t$, a pair of elements $Y^t_\pm$
which respectively belong to the two Borel subalgebras $\CB^t_\pm$ of
$\CH_{taf}$:
\debut
\tilde \pi ~:\quad Y^s \in \CK^s_{af} \to (Y^t_+,Y^t_-) \in \CB^t_+\times \CB^t_- 
\label{doubling}
\fin
We add the superscript ${}^t$ in order to recall that these 
algebras are linked with the $t$ parametrization.
$Y^t_\pm$ are semi infinite formal power series in $t$, they thus have
to be thought as elements of an appropriate completion of $\CB^t_\pm$. In their
domains of convergence, which are never overlapping, they coincide 
with the same analytic function, namely $Y^s$.
Taking $t$ on the unit circle, the differences $Y^t_+-Y^t_-$ are linear 
combinations of derivatives of Dirac $\de$-functions.
Thus $Y^t_\pm$ are solutions of singular Riemann-Hilbert problems.
The map $\tilde \pi$ is not a Lie algebra homorphism 
from $\CK^s_{af}$ to the father Lie algebra $\CH_{taf}$. 
However, it is a Lie algebra homomorphism from $\CK^s_{af}$ to the
dressing algebra in which the $\pm$ components commute separately, cf.
eq.(\ref{comm2}). Indeed,
\debut
\tilde \pi\({ \[{Y^s_1, Y^s_2}\]}\) = 
\({\[{Y^t_1,Y^t_2}\]_+, \[{Y^t_1,Y^t_2}\]_- }\)=
\({\[{Y^t_{1\,+},Y^t_{2\,+}}\], \[{Y^t_{1\,-},Y^t_{2\,-}}\] }\) 
\label{homo}
\fin

The comparison between dressing and Geroch transformations can then be 
made more explicit. First, as we have just seen, we have to consider wave 
functions $\hat \psi_G$ which are in the untwisted affine Kac-Moody group $G^s_{af}$.
Since the Lax connection (\ref{Lax}) at $k=0$ belongs to the subalgebra $V \CH$
with the central element set to zero, the corresponding $\Psi_G$ can be written as
\debut
\Psi_G = \Psi_H\cdot n,\quad{\rm with}\quad \Psi_H\in {\bf VH},~~ n\in G^s_{af}
\label{bigwave}
\fin
where $\Psi_H$ is solution of the linear system, and
$n$, independent of the space-time position, will be
called the ``normalizing factor".
Since by definition of $\bf VH$, $\tau'(\Psi_H)=\Psi_H$, $n$ is
related to the so-called monodromy $m$:
\debut
m \equiv \Psi_G^{-1}\cdot \tau'(\Psi_G) = n^{-1}\cdot \tau'(n) 
\label{mono}
\fin
Of course $n$ is defined up to left multiplications by elements
in $K^s_{af}$. It belongs to the coset space $K^s_{af}\bacs G^s_{af}$,
and we have the equivalence $(\Psi_H,n)\sim (\Psi_H g^{-1}, g n)$
for $g\in K^s_{af}$. Let us choose a gauge for $n$, by 
imposing $\Psi_H(x_0)=1$ at some point $x_0$. Then, since $n=\Psi_G(x_0)$
the triangularity condition (\ref{condi}) forces $n$ to be
regular at $s=0$. We then have a usual action of $G^s_{af}$
on $K^s_{af}\bacs G^s_{af}$ by 
\debut
n ~\to  ~ M_{e_s}(n) = \kappa^s(n,e_s)\cdot n\cdot e_s^{-1},\quad {\rm with}\quad
\kappa^s(n,e_s) \in K^s_{af},\quad e_s\in G^s_{af} \label{msurn}
\fin
where $\kappa^s(n,e_s)$ is determined by requiring that 
$M_{e_s}(n)$ fulfills the gauge condition.

Let us now consider the Geroch transformation (\ref{glob}).
By considering it at the point $x_0$ where $\Psi_H$ is normalized to one,
it induces an action on $n$ which by construction coincides with the action 
(\ref{msurn}). Thus, the Geroch transformations (\ref{glob})
may be written as:
\debut
\Psi_G &\to& \({K^s\cdot \Psi_H\cdot \kappa^{s\,-1}}\)\cdot M_{e_s}(n)
\quad,\quad {\rm with}~~\kappa^s= K^s(x_0) 
\label{glob2}\\
&\to& \({ K^t_\pm\cdot \Psi_H\cdot \kappa^{t~-1}_\pm}\) \cdot M_{e_s}(n)
\quad,\quad {\rm with}~~ (K^t_+,K^t_-)=\tilde \pi(K^s) \non
\fin
In the last line we used the homorphism $\tilde \pi$ to identify
$K^s$ with its images in $\exp\CB_\pm^t$. 
We then recognize the dressing transformations.

\bigskip
{\bf To summarize}: we thus have argued that the Geroch transformations
can be understood as combinations of dressing transformations
and actions (\ref{msurn}) on the ``normalizing factor" $n$. 
Recall that we made contact with the  dressing transformations
by dropping the central charge $k$. 
This identification is valid only once
the wave function has been lifted to the untwisted
affine Kac-Moody $SL(2,\Rmath)$ group by imposing the
triangular gauge (\ref{condi}).

However, a more mathematically rigorous identification of Geroch transformations
with dressing transformations would require a better construction of the map between
the two loop groups $G^s_{af}$ and $G^t_{af}$.

\bigskip

\subsection{The phase space contains only one dressing orbit.}
We now present arguments indicating that the
vacuum orbit under the group of dressing transformations
covers most of the phase space. In other words, any solution
of truly 2D reduced Einstein's gravity should be obtainable 
by dressing from the vacuum solution. 

Let us make more precise what phase space we consider.
It is defined as the collection of gauge inequivalent solutions
of the dimensionaly reduced Einstein's equations (\ref{EL1},\ref{EL2},
\ref{EL3}) and (\ref{EL4}) for a fixed and given harmonic function $\rho$
such that $\d_\pm\rho\not=0$.
In particular we insist that $\hat \sig$ is solution
of the first order equation.

Let us consider a solution of Einstein's equations (\ref{EL1},\ref{EL2},
\ref{EL3}) and (\ref{EL4}). We choose the gauge $Q_-=0$ and
parametrize $Q_+$ as follows:
\debut
Q_-=0\quad,\quad Q_+=h(0)\d_+h(0)^{-1} \label{gaga}
\fin
for some $h(0)\in H$.
Consider the Lax connexion $A_\pm$ and its wave function $\Psi$
which we choose in $\bf V H$. 

The proof of the fact that there is only one dressing orbit in
the phase space consists in analysing the form of the wave functions.
As we are going to argue in the following, assume that we may decompose
any solution $\Psi$ of the linear system (\ref{linear})
in the two following ways up to right multiplications by constant 
group elements:
\debut
\Psi &=& e^{-\zeta \frac{k}{2}}\cdot \psi_- \Psi_0 \quad
{\rm with}\quad \psi_-\in \exp\CB_- 
\label{comp1}\\
\Psi &=& e^{\zeta \frac{k}{2}}\cdot h(0)\psi_+ \Psi_0 \quad
{\rm with}\quad \psi_+\in \exp\CB_+ 
\label{comp2}
\fin
where $\Psi_0$ is the vacuum wave function.
We are now going to show that assuming eqs.(\ref{comp1},\ref{comp2})
to be true implies that $\Psi$ is in the dressing orbit of the vacuum.
With the gauge choice (\ref{gaga}), the compatibility of these
decompositions tells us that $\psi_\pm$ are actually exponentials of
elements of degree strictly positive (negative).
For a given solution $Q_\pm,~P_\pm,~\hat \sig$, the two  assumed decompositions
(\ref{comp1},\ref{comp2}) are by hypothesis solutions of the same linear 
system. Therefore they differ by a right multiplication. 
In other words, there exists $g\in H_{taf}$ such that:
\debut
e^{-\zeta \frac{k}{2}}\cdot \psi_- \Psi_0\cdot g
= e^{\zeta \frac{k}{2}}\cdot h(0) \psi_+ \Psi_0 \non
\fin
Alternatively,
\debut
e^{2\zeta \frac{k}{2}}\cdot \psi_-^{-1} h(0)\psi_+
= \Psi_0 g\Psi_0^{-1} \label{orbi1}
\fin
Since $\psi_\pm \in \exp\CB_\pm$, we recognize in eq.(\ref{orbi1})
the solution of the problem of factorizing $(\Psi_0 g\Psi_0^{-1})$.
In other words, eq.(\ref{orbi1}) tells us that if the two decompositions
(\ref{comp1}) and (\ref{comp2}) hold then the $\pm$ - components of
$(\Psi_0 g\Psi_0^{-1})$ are:
\debut
e^{\pm\zeta \frac{k}{2}}\cdot h_\pm(0) \psi_\pm 
= (\Psi_0 g\Psi_0^{-1})_\pm \label{orbi2}
\fin
with $h_-(0)=1$ and $h_+(0)=h(0)$. Comparing the assumed
decompositions eqs.(\ref{comp1},\ref{comp2})
for the wave function with eq.(\ref{orbi2}) implies that
the wave function is in the dressing orbit of the vacuum since:
\debut
\Psi = e^{-\zeta \frac{k}{2}}\cdot \psi_- \Psi_0 g_-^{-1}
=e^{\zeta \frac{k}{2}}\cdot h(0)\psi_+ \Psi_0 g_+^{-1}
= D_{(g_-,g_+)}(\Psi_0) \label{orbi3}
\fin
where $g_\pm$ are the $\pm$ - components of g, ie. $g=g_-^{-1}g_+$.

The arguments described below will show that if the wave function $\Psi$
admits the decompositions (\ref{comp1}) or (\ref{comp2}) then $\zeta$ is 
equal to the conformal factor $\hat \sig$,
\debut
\zeta=\hat \sig. \label{1order}
\fin
In other words, imposing the condition that $\Psi\cdot \Psi_0^{-1}
\in \exp\CB_-$ demands that $\zeta=\hat \sig$ is solution
of the first order equation (\ref{EL4}).
In particular, the wave functions in the dressing orbit of the
vacuum are of the form (\ref{comp1}) or (\ref{comp2}). Thus
the conformal factors $\hat \sig$ obtained by dressing the vacuum
are solutions of the first order eq.(\ref{EL4}).

\bigskip

This discussion shows that proving that there is only one
orbit in the phase space is equivalent to prove the 
decompositions (\ref{comp1},\ref{comp2}).
Let us now argue that these are indeed correct.
Let $\Psi$ be a solution of the linear system.
Decompose $\Psi$ as $\Psi=\hat \psi \cdot \Psi_0$ where $\Psi_0$
is the vacuum wave function. Then, as in eq.(\ref{akmbis}),
$\hat \psi$ is a solution of the following equation:
\debut
(\hat D_\pm\hat \psi)\cdot \hat \psi^{-1}  
+ Q_\pm + P_\pm \otimes t^{\pm 1} 
\mp (\d_\pm \hat \sig) \frac{k}{2} = 0 
\label{linbis}
\fin
where $\hat D_\pm\hat \psi = \d_\pm\hat \psi \pm d_\pm [E_\pm, \hat \psi]$.
To prove the decomposition (\ref{comp1}) one first has to prove
that we may choose $\hat \psi$ in $\exp(\CB_- \oplus \Cmath k)$.
But, since in the Lax connection there are elements of positive
degree,  the linear system (\ref{linbis}) is not obviously
a linear system in $(\CB_- \oplus \Cmath k)$.
The point which allows to prove that we may choose $\hat \psi$ 
in $\exp(\CB_- \oplus \Cmath k)$, is that we may compensate the
term $P_+\otimes t$ which is of degree one by the term of degree one in
the commutator $[E_+, \hat \psi]\hat \psi^{-1}$. Since commuting 
with $E_+$ shifts the degree by two, this condition fixes the
degree minus one component of $\hat \psi$. More precisely,
let us decompose $\hat \psi$ as:
\debut
\hat \psi = e^{-\zeta\frac{k}{2}} \cdot h_{(-1)}\cdot M 
\quad {\rm with} \quad h_{(-1)} = e^{V_{(-1)}\otimes t^{-1}} 
\label{form1}
\fin
with $M$ exponential of elements of degree less than minus two.
Plugging this ansatz into eq.(\ref{linbis}) and requiring the
cancellation of the terms of degree one imposes to choose
$V_{(-1)}$ such that:
\debut
d_+~V_{(-1)} =2 P_+ \non
\fin
Requiring the cancellation of the terms of degree zero demands:
\debut
\d_-\zeta= \d_-\hat \sigma \quad , \quad 
\d_+\zeta = -\d_+\hat \sig + \inv{4} d_+ tr(V_{(-1)}^2) \non
\fin
This means $\zeta=\hat \sig$. Then, the terms of degree minus
one in (\ref{linbis}) automatically cancel and the factor $M$
in (\ref{form1}) is solution of:
\debut
(\hat D_+ M)M^{-1} + h_{(-1)}^{-1}h(0)\d_+(h(0)^{-1}h_{(-1)}) + P_+\otimes t^{-1} =0 
\label{ext1}\\
(\hat D_- M)M^{-1} + \inv{2}d_- V_{(-1)}\otimes t^{-3}
+ h_{(-1)}^{-1}\bigl({ \d_-h_{(-1)}\cdot h_{(-1)}^{-1} 
- \d_-V_{(-1)}\otimes t^{-1}}\bigr) h_{(-1)}=0
\label{ext2}
\fin
These are two equations compatible with $M$ being exponential
of elements of degree less than minus two. Thus it is an equation
in the Borel subalgebra of the loop algebra, ie. no new central term can
be produced, they all have already been determined. 
The equation (\ref{ext1}) along the $z_+$ direction determines
algebraically $M$ degree by degree. This follows from the 
fact that this equation contains $E_+$ which increases the degree by two.
Hence we have parallely transported $\Psi$ 
along the $z_+$-direction starting from a given normalization point $x_0$ 
at which $\Psi(x_0)=1$ such that $\hat \psi \in \exp(\CB_- \oplus \Cmath k)$. 
We may then transport it
along the $z_-$-direction to arrive at any point $(z_+,z_-)$.
Since the component $A_-$ of the Lax connexion  is of negative degree,
the resulting wave function will satisfy the decomposition (\ref{comp1}).
A similar proof applies to the decomposition (\ref{comp2}).
\bigskip

\section{Dressing orbit of the vacuum and $\tau$-functions.}

We now describe the solutions obtained by dressing the vacuum solution.
There are two ways to compute the metrics depending whether one is willing to
compute the original fields $\De,~N$ or their duals $\De^*,~N^*$. Ie:

(i) As it is clear from the formula (\ref{dresq},\ref{dresp}) and
(\ref{drsig}), in order to evaluate the connexion $(Q_\pm+P_\pm)$
for the dressed solutions one needs
to compute $\zeta$, $h(0)$ and $X_{(\pm 1)}$. 
This will be done by introducing $\tau$-functions.
This method apply to the $H\bacs G$ models.

(ii) In the gravity case, we can use the specific form of
the elements (\ref{y*pm}) generating the Ehlers transformation
to directly compute the dual fields $\De^*,~N^*$ and $\hat \sig^*$.

\subsection{$\tau$-functions in the $H\bacs G$ case.}
Let us first describe how to compute $\zeta$ and $h(0)$.
Let ${\La}$ be a highest
weight for $V\CH$, and $\ket{\la_1}$ and $\ket{\la_2}$ be two
vectors of the highest representation of $\La$.
We define $\tau$-functions $\tau_0^{\la_1;\la_2}(z_+,z_-)$ by:
\debut
\tau_0^{\la_1;\la_2}(z_+,z_-) = \bra{\la_1} h(0)\ket{\la_2}~ e^{\zeta k}
\label{deftau}
\fin
It depends on the element $g=g_-^{-1}g_+$ of the dressing group.
Alternatively, we may view $\tau_0^{\la_1;\la_2}$ as a matrix 
$\tau_0$ such that $\bra{\la_1}\tau_0\ket{\la_2}
=\tau_0^{\la_1;\la_2}$, Ie.
\debut
\tau_0 = h(0)~\exp(\zeta k) \label{deftau2}
\fin
It can be evaluated by computing
the expectation value of $(\Psi_0 g\Psi_0^{-1})$.
Indeed, if $\la_1$ and $\la_2$ are two
vectors of the highest representation of $\La$,
then by definition of $h(0)$ and $\zeta$, eq.(\ref{hzero}), we have:
\debut
\bra{\la_1}(\Psi_0 g\Psi_0^{-1})\ket{\la_2} &=&
\bra{\la_1} h_-(0)^{-1}h_+(0)\ket{\la_2}~ e^{\zeta k}
=\bra{\la_1} h(0)\ket{\la_2}~ e^{\zeta k} 
\label{vevtau}
\fin
Thus, given the explicit expressions for the 
vacuum wave function, eq.(\ref{psivac1},\ref{psivac2}), we obtain:
\debut
\tau_0^{\la_1;\la_2}(z_+,z_-)
&=&\bra{\la_1} (\Psi_0 g\Psi_0^{-1})\ket{\la_2 } \non\\
&=& \({\frac{\rho^2}{AB}}\)^{h_\La} \cdot  
\bra{\la_1} \({A(z_+)}\)^{-E_+} \cdot g \cdot
\({B(z_-)}\)^{-E_-} \ket{\la_2} \label{theta0} 
\fin
with 
\debut
A(z_+)= a(z_+)+\inv{2} \quad,\quad
B(z_-)= b(z_-)+\inv{2} \label{defAetB}
\fin
These variables have been introduced in order to simplify the
expression of the vacuum wave function:
$\Psi_0= (\frac{B}{\rho})^{E_+} B^{E_-}= (\frac{\rho}{A})^{E_-}A^{-E_+}$.

Next we have to compute $X_{(\pm 1)}$.
This can be done by inserting elements of degree plus or minus one
inside the expectation value (\ref{vevtau}). For
any vectors $\ket{\la_1}$ and $\ket{\la_2}$ of the highest representation $\La$,
and for any $Y\in {\bf r}$, we define two new $\tau$-functions
$\tau_{(\pm 1);Y}^{\la_1;\la_2}(z_+,z_-)$ by:
\debut
\tau_{(+1);Y}^{\la_1;\la_2}(z_+,z_-) &=& 
\bra{\la_1}(\Psi_0 g \Psi^{-1}_0)\cdot(Y\otimes t^{-1})\ket{\la_2}
\label{tauy1}\\
\tau_{(-1);Y}^{\la_1;\la_2}(z_+,z_-) &=& 
\bra{\la_1}(Y\otimes t)\cdot(\Psi_0 g \Psi^{-1}_0)\ket{\la_2}
\label{tauy2}
\fin
Inserting the explicit expression (\ref{psizero})
of the vacuum wave function, we get explicit expressions
for $\tau_{(\pm 1);Y}^{\la_1;\la_2}$:
\debut
\tau_{(+1);Y}^{\la_1;\la_2}&=& 
\({\frac{\rho^2}{AB}}\)^{h_\La}
\({\frac{\rho}{B}}\)^\half\cdot 
\bra{\la_1} \({A(z_+)}\)^{-E_+} \cdot g \cdot
\({B(z_-)}\)^{-E_-} (Y\otimes t^{-1})\ket{\la_2} \label{theta1}\\
\tau_{(-1);Y}^{\la_1;\la_2}&=&  
\({\frac{\rho^2}{AB}}\)^{h_\La} 
\({\frac{\rho}{A}}\)^\half\cdot 
\bra{\la_1}(Y\otimes t) \({A(z_+)}\)^{-E_+} \cdot g \cdot
\({B(z_-)}\)^{-E_-} \ket{\la_2} \label{theta2}
\fin
As before, we may view $\tau_{(\pm 1);Y}^{\la_1;\la_2}$ as
a matrix $\tau_{(\pm 1);Y}$ such that 
$\bra{\la_1} \tau_{(\pm 1);Y}\ket{\la_2}
=\tau_{(\pm 1);Y}^{\la_1;\la_2}$. Moreover, using the decomposition
(\ref{hzero}) of $(\Psi_0g\Psi_0^{-1})$ which defines $X_{(\pm 1)}$, 
one may relate $\tau_{(\pm 1);Y}$ to $X_{(\pm 1)}$:
\debut
\tau_{(+ 1);Y} &=&
\tau_0\cdot\({[X_{(+1)},Y] + tr(X_{(+1)}Y)\frac{k}{2}}\)
\label{tauy1bis}\\
\tau_{(- 1);Y} &=&
\({[X_{(-1)},Y] - tr(X_{(-1)}Y)\frac{k}{2}}\)\cdot \tau_0
\label{tauy2bis}
\fin
Hence, given $ \tau_0$ and $ \tau_{(\pm 1);Y}$
we can extract $\zeta$, $h(0)$ and $X_{(\pm 1)}$.
\bigskip

{\bf To summarize}: the three $\tau$-matrices $\tau_0$
and $\tau_{(\pm 1);Y}$
encode all local informations concerning the dressed solutions.
All the fields, $Q_\pm$, $P_\pm$ and $\hat \sig$, or
alternatively $\zeta$, $h(0)$ and $X_{(\pm 1)}$ can be
reconstructed from them using the relations (\ref{deftau2})
and (\ref{tauy1bis},\ref{tauy2bis}). 
The solutions obtained by dressing the vacuum solution are thus 
determined by the three expectation values 
(\ref{theta0},\ref{theta1},\ref{theta2}).

This system being integrable, there exists an infinite hierarchy of 
equations of motion begining with the Ernst equations. We have
seen that these equations are linked to the elements $E_\pm$. Thus 
computing explicitly the commuting flows will amount to constructing two
abelian infinite dimensional subalgebras of 
$V\CH= Vir \times \CH_{af}$ containing $E_\pm$.

\subsection{$\tau$-functions for 2D reduced gravity.}
We now consider the gravity case $SO(2)\bacs SL(2,\Rmath)$.
Since $H=SO(2)$ is abelian, the $\tau$-functions
are really functions not matrices. Recall that
the affine algebra is then isomorphic to the affine
algebra $A_1^{(1)}$ in the principal gradation.
Recall also the definition of the highest weight vectors $\ket{\La_\pm}$
given in eq.(\ref{lapm}).

We first define the $\tau$-functions $\tau_0$ associated to $\ket{\La_\pm}$
which allows us to compute $h(0)$ and $\zeta$:
\debut
\tau^\pm_0(z_+,z_-) \equiv \bra{\La_\pm}
(\Psi_0 g \Psi_0^{-1})\ket{\La_\pm} 
= \exp(\zeta \mp \frac{i}{4}\varphi ) 
\label{tau2}
\fin
where we parametrized $h(0)$ by $h(0)=\exp\({-\frac{\varphi}{2} (\sig^+-\sig^-)}\)$.

Next we define the $\tau$-functions $\tau_{(\pm 1);Y}^\pm$ which allows
us to compute $X_{(\pm 1)}$. It turns that it is enough to only
consider the functions $\tau_{(\pm 1);Y}^\pm$ with $Y=\sig^z$.
Parametrizing $X_{(\pm 1)}$ by
$X_{(\pm 1)}=X_{(\pm 1)}^1 \sig^z + X_{(\pm 1)}^2 (\sig^++\sig^-)$
and using eq.(\ref{tauy1bis},\ref{tauy2bis}), we find~:
\debut
\tau_{(+1)}^\pm(z_+,z_-) &\equiv& \bra{\La_\pm}
(\Psi_0 g \Psi_0^{-1})\cdot(\sig^z\otimes t^{-1})\ket{\La_\pm}
=~\tau_0^\pm \cdot \({ X^1_{(+1)} \mp i X^2_{(+1)} }\) \label{tauy1su2}\\
\tau_{(-1)}^\pm(z_+,z_-) &\equiv& \bra{\La_\pm}  (\sig^z\otimes t)\cdot
(\Psi_0 g \Psi_0^{-1})\ket{\La_\pm} 
=-~ \({ X^1_{(-1)} \mp i X^2_{(-1)} }\)\cdot \tau^\pm_0 \label{tauy2su2}
\fin
Clearly we may introduce generating functions,  depending on an
infinite number of variables $x, \bar x$, for these $\tau$-functions
by defining $\tau_\pm(x,\bar x)$ as:
\debut
\tau_\pm(x,\bar x) \equiv \bra{\La_\pm} e^{H(x)}\cdot
(\Psi_0 g \Psi_0^{-1})\cdot e^{\bar H(\bar x)}\ket{\La_\pm}
\fin
with $H(x)=\sum_{n odd} (\sig^z\otimes t^n)x_n$ and similarly for 
$\bar H(\bar x)$.
The $\tau$-functions $\tau^\pm_0$, $\tau^\pm_{(\pm1)}$ are gauge
invariant.

Recall that given these $\tau$-functions, 
the dressed connexion $(Q_\pm+P_\pm)$ and the
dressed value of $\hat \sig$ are computed using eq.(\ref{drsig})
and (\ref{drq},\ref{drp}). Since we are actually dressing
the vacuum solution for which $Q_\pm=P_\pm=\hat \sig=0$, we have:
\debut
P_\pm &=& \frac{d_\pm}{2} 
\({ X_{(\mp 1)}^1~\cos\varphi_\mp- X_{(\mp 1)}^2~\sin\varphi_\mp }\)\cdot \sig^z
+\frac{d_\pm}{2}\({ X_{(\mp 1)}^1~\sin\varphi_\mp
+ X_{(\mp 1)}^2~\cos\varphi_\mp }\)\cdot (\sig^++\sig^-) \non\\
Q_\pm &=& \half (\d_\pm \varphi_\pm)\cdot (\sig^+-\sig^-) 
\label{drvac}\\
\varphi &=& \varphi_+ - \varphi_- \non\\
\hat \sig &=& \zeta \non
\fin
where $d_\pm = \rho^{-1}\d_\pm\rho$.
Different choices of $\varphi_\pm$ correspond to different gauge choices.
Using eq.(\ref{drpbis}) gives an alternative but equivalent expression
for $P_\pm$:
\debut
P_\pm &=& -\rho^\half\({\d_\pm(\rho^{-\half} X_{(\pm 1)}^1)~\cos\varphi_\pm
- \d_\pm(\rho^{-\half}X_{(\pm 1)}^2)~\sin\varphi_\pm }\)\cdot \sig^z \label{drvac2}\\
&~&-\rho^\half\({ \d_\pm(\rho^{-\half} X_{(\pm 1)}^1)~\sin\varphi_\pm
+ \d_\pm(\rho^{-\half}X_{(\pm 1)}^2)~\cos\varphi_\pm }\)
\cdot (\sig^++\sig^-) \non
\fin
Comparing eqs.(\ref{drvac}) and (\ref{drvac2}) gives algebraic
relations involving the $\tau$-functions and their derivatives.
Note that the field $\hat \sig$ obtained by dressing is a solution
of the first order eq.(\ref{EL4}). It is is given by the product of the 
$\tau$-functions:
\debut
 \tau_0^+\tau^-_0=e^{2\hat \sig} \quad ,\quad 
\frac{\tau^+_0}{\tau^-_0}= e^{-\frac{i}{2}\varphi} \label{tautau}
\fin
A $\tau$-function related to the conformal factor $\hat \sig$ has
been introduced in ref.\cite{koni} using isomonodromic deformation.
\bigskip

\subsection{Algebraic evaluation of the dual fields.}
We are now going to describe how to use the particular form
of the elements (\ref{y*pm}) generating the Ehlers transformation
to compute the dual fields $N^*,~\De^*$. The important point is that these
elements have very simple expressions in terms of the dual fields.

Let $\hat \CE^*=\hat \CE^*_+-\hat \CE^*_-$ with
 $\hat \CE^*_\pm \in \CB_\pm$ be defined by:
\debut
\hat \CE^*_\pm= \pm \[{ (\sig^+-\sig^-)\otimes
\({\frac{1+t^{\pm 2}}{1-t^{\pm 2}}}\) - (\sig^++\sig^-)\otimes
\({\frac{2t^{\pm 1}}{1-t^{\pm 2}}}\) }\]
\label{defx*}
\fin
The elements $\hat Y^*_\pm$ generating the dressing transformations
corresponding to the Ehlers transformation can then be expressed
in a very simple way, cf eq.(\ref{y*pm}):
\debut
\hat Y^*_\pm = \pm N^*~ \frac{k}{2} + \De^{*~-1}~ \hat \CE^*_\pm
\label{y*pmbis}
\fin

Recall now the fact that the elements $\hat Y^*_\pm$ are generating the
Ehlers transformation means that
$\d_\nu \hat Y^*_\pm + [A_\nu, \hat Y^*_\pm]= - J^*_+(A_\nu)$,
where $A_\nu$ is the Lax connexion in the triangular gauge (\ref{QPtri})
and $J^*(A_\nu)$ its Ehlers transform (\ref{j*a}).
This implies that the difference $\hat Y^*=\hat Y^*_+-\hat Y^*_-$
satisfy: $ \d_\nu \hat Y^* + [A_\nu, \hat Y^*]= 0$. Thus,
\debut
\hat Y^* = N^*~ k + \inv{\De^*}~ \hat \CE^* = \Psi_{tr} \hat \CE^* \Psi_{tr}^{-1}
\label{magic}
\fin
where $\Psi_{tr}\in {\bf VH}$ is the wave function,
solution of the linear system in the triangular gauge (\ref{QPtri})
and normalized such that $\Psi_{tr}(x_0)=1$ at some point $x_0$.
For example we may choose the point $A=B=1$.
Notice that dilatating  $\hat \CE^*$ or translating it by $k$ amount to
implement Ehlers transformations on $N^*, ~\De^*$.

The relation (\ref{magic}) may be interpreted more algebraically.
Indeed consider its factorization in $\pm$ components:
\debut
\hat Y^*_\pm \equiv \pm N^* \frac{k}{2} + \inv{\De^*} \hat \CE^*_\pm
= (\Psi_{tr} \hat \CE^* \Psi_{tr}^{-1})_\pm \label{orb2}
\fin
Since $(\hat \CE^*_+,\hat \CE^*_-)$ is an element of the dressing algebra,
the map $\hat \CE^*_\pm\to (\Psi_{tr} \hat \CE^* \Psi_{tr}^{-1})_\pm$  with
$\Psi_{tr}$ varying as a function of $(z_+,z_-)$ induces an orbit on the dressing
algebra. This orbit is two dimensional and
the dual fields $N^*,~\De^{*~-1}$, which are the components
of $\hat Y^*_\pm$  on $k$ and $\hat \CE^*_\pm$, are simply the coordinates
of this orbit.

Taking expectation values of the magic formula (\ref{magic}) allows us to evaluate
$N^*,~\De^*$. Indeed the defining properties of the two highest
weight vectors $\ket{\La_\pm}$ implies that $\bra{\La_\pm}k\ket{\La_\pm}=1$
and $\bra{\La_\pm}\hat \CE^*\ket{\La_\pm}=\pm i$. Thus:
\debut   
N^*~ \pm~ \frac{i}{\De^*} = \bra{\La_\pm}\Psi_{tr} \hat \CE^* \Psi_{tr}^{-1}\ket{\La_\pm}
\label{magic2}
\fin
In particular for wave functions $\Psi_{tr}$  which are in the dressing orbit
of the vacuum wave function $\Psi_0$, we have $\Psi_{tr}= \Th_\pm\cdot\Psi_0\cdot
g_\pm^{-1}$. Demanding that $\Psi_{tr}$ be the wave function in the
triangular gauge fixes the arbitrarines in the factorization problem.
Hence, the expectation value (\ref{magic2}) becomes:
\debut
N^*~ \pm~\frac{i}{\De^*} =
\frac{\bra{\La_\pm}\Psi_0\cdot (g_-^{-1} \hat \CE^* g_+)\cdot \Psi_0^{-1}\ket{\La_\pm}}{
\bra{\La_\pm}\Psi_0\cdot (g_-^{-1}g_+)\cdot \Psi_0^{-1}\ket{\La_\pm}}
\label{magic3}
\fin
A $SO(2)$ gauge transformation transformation preserving the normalization
$\Psi(x_0)=1$ corresponds to the gauge transformation $g_\pm \to hg_\pm$
for a constant $h\in SO(2)$. Under such transformation the numerator
of (\ref{magic3}) transforms as:
\debut
\bra{\La_\pm}\Psi_0\cdot (g_-^{-1} \hat \CE^* g_+)\cdot \Psi_0^{-1}\ket{\La_\pm}
\to
\bra{\La_\pm}\Psi_0\cdot (g_-^{-1} h^{-1}\hat \CE^*h g_+)\cdot \Psi_0^{-1}\ket{\La_\pm}
\label{hXh}
\fin
Imposing the triangular gauge (\ref{QPtri}) will require fixing this
parameter $h$.
 
In the denominator of eq.(\ref{magic3})
we recognize the two $\tau$-functions $\tau^\pm_0$
introduced above. As we have explained, their product directly give
$\hat \sig$ by $\tau^+_0\tau^-_0 = e^{2\hat \sig}$. Using the
duality relation (\ref{dual3}) this gives:
\debut
\rho^\half \De^*~ e^{2\hat \sig^*} = \tau^+_0\tau^-_0
\label{dualtau}
\fin
Formula (\ref{magic3}) and (\ref{dualtau}) are those quoted in the
introduction.
\bigskip
 
{\bf To summarize:} The ``Ehlers" potential $\De^{*~-1} + i N^*$ and
the conformal factor $e^{2\hat \sig^*}$ can be algebraically and
directly evaluated in terms of the expectation values (\ref{magic2})
or (\ref{magic3}) and (\ref{dualtau}).
In other words, the formula (\ref{magic3}) and (\ref{dualtau})
give an algebraic formula of the dual metric (\ref{metric4}).

\section{Solutions and vertex operators.}
We now describe how to use vertex operators to
compute the $\tau$-functions.
We consider the gravity case $SO(2)\bacs SL(2,\Rmath)$,
the computations of the $\tau$-functions in the
$SO(n)\bacs SL(n,\Rmath)$ case is similarly simple.
\subsection{Vertex operator representations.}
These $\tau$-functions can be computed using vertex operators
since level one representations of $A_1^{(1)}$ in the
principal gradation can be constructed using a $\Zmath_2$ twisted
free bosonic field \cite{lewi}.
 
Let us denote by $Z(\mu)$ this field:
\debut
Z(\mu)=-i\sum_{n~odd}~p_{-n}\frac{\mu^n}{n}\quad {\rm with}\quad
[p_n,p_m]=n\de_{n+m,0} \non
\fin
The operators $p_n$ generate a Fock space,
we denote by $\ket{0}$ its vacuum: $p_n\ket{0}=0$ for $n>0$.
For any complex number $u$, let $W_u(\mu)$ be the vertex operators:
\debut
W_u(\mu)= :\exp(-iu Z(\mu)): \label{verop}
\fin
Their $N$-point functions are:
\debut
\vev{\prod_p W_{u_p}(\mu_p)} =
\prod_{p<q}\({\frac{\mu_p-\mu_q}{\mu_p+\mu_q}}\)^{u_p\cdot u_q/2}\label{wick}
\fin
The level one representations with highest weight $\La_\pm$ are
defined by:
\debut
i\mu \frac{dZ(\mu)}{d\mu} &=& \sum_{n~odd}
(\sig^z\otimes t^n) \mu^{-n} \label{repv}\\
\pm i~ W_2(\mu) &=& 2 \sum_{n~even}
((\sig^+-\sig^-)\otimes t^n)\mu^{-n}
-2\sum_{n~odd}((\sig^++\sig^-)\otimes t^n)\mu^{-n} \non
\fin
The highest weight vectors $\ket{\La_\pm}$ are identified with the
vacuum vector $\ket{0}$. Note that the two representations $\La_\pm$
only differ by the sign in front of the vertex operator $W_2(\mu)$.
The Virasoro algebra acts on the Fock space generated by the $p_n$.
The Virasoro generators are represented by
$$\sum_n (L_n-\inv{16}\de_{n,0})
 \mu^{-2n-2} = -\inv{4}: (\d_\mu Z)^2:$$
The conformal weights are $h_{\La_\pm}=\inv{16}$.
\bigskip
 
Eq.(\ref{repv}) means that $(\sig^z\otimes t^n)$ is
represented by $p_n$. Therefore, we may represent the vertex operators
(\ref{verop}) as:
\debut
W_u(\mu)= \exp\({-\bu\otimes\log\({\frac{1+\mu/t}{1-\mu/t}}\)}\)\cdot
\exp\({\bu\otimes\log\({\frac{1+t/\mu}{1-t/\mu}}\)}\) \label{wuh}
\fin
with $\bu = \frac{u}{2}\sig^z$. Thus they have a natural interpretation
in the affine group $H_{taf}$.
\bigskip
 
Since $\pm W_2(\mu)$ are generating functions representing
elements of the affine algebra, and since they are nilpotent, ie.
$W_2(\mu)W_2(\mu)=0$ inside any correlation functions,
the products
\debut
\prod_j\({ 1 \pm y_j W_2(\mu_j)}\) \label{gsol}
\fin
are representations in $\La_\pm$ of elements of the complex Kac-Moody group.
As shown in ref.\cite{babe2} these products are representations
of products of elements  $g_-^{-1}g_+\in H_{taf}$ with $g_\pm$ exponentials
of elements in the Lie algebra $\CH_{taf}$.
More precisely, let $g_\pm(j)$ be the elements defined by
\debut
g_\pm(j)= e^{\pm r_j\frac{k}{2}}~e^{v_j(\sig^+-\sig^-)/2}~
e^{s_j \hat \CE^*_\pm(\mu_j)/2} \label{defg-g+}
\fin
with
\debut
\hat \CE^*_\pm(\mu)= \pm \[{
(\sig^+-\sig^-)\otimes\({\frac{1+(t/\mu)^{\pm2}}{1-(t/\mu)^{\pm2}}}\)
-(\sig^++\sig^-)\otimes\({\frac{2(t/\mu)^{\pm1}}{1-(t/\mu)^{\pm2}}}\) }\]
\label{defxmu}
\fin
Then \cite{babe2},
\debut
g_-^{-1}(1)\cdots g_-^{-1}(M)\cdot g_+(M)\cdots g_+(1)=
\prod_{j=1}^M\({ 1 \pm y_j W_2(\mu_j)}\) \label{g-g+w}
\fin
Eq.(\ref{g-g+w}) is valid in the representation $\La_\pm$.
The relation between the parameters $(s_j,r_j,v_j)$ and $(y_j,\mu_j)$
is explained in Appendix E. Notice that $\hat \CE^*_\pm(\mu=1)=\hat \CE^*_\pm$
as defined in eq.(\ref{defx*}).

\subsection{Explicit expressions for the $\tau$-functions.}
We may in principle consider the $\tau$-functions for
any dressing element, but in order to be able to compute
them using the vertex operator representation, one should
consider elements made of products of elements of the form
(\ref{gsol}) or (\ref{wuh}); ie.
\debut
g=g_-^{-1}g_+
=\prod_pW_{u_p}(\mu_p) \prod_j\({ 1 \pm y_j W_2(\mu_j) }\) \label{gvertex}
\fin
Here $y_j$, $\mu_j$ and $\mu_p$ are parameters.
Real solutions to Einstein's gravity correspond to $y_j$ imaginary
and $\mu_j$ real.
 
Using eqs.(\ref{theta0},\ref{theta1},\ref{theta2}) and
 the fact that $(\sig^z\otimes t^{\pm1})$ are
represented by $p_{\pm 1}$ in the level one representation
$\La_\pm$, we find the following expressions for the $\tau$-functions:
\debut   
\tau^\pm_0 &=&  \vev{\Psi_0\cdot\prod_pW_{u_p}(\mu_p)
\prod_j\({ 1 \pm y_j W_2(\mu_j)}\)\cdot \Psi_0^{-1} }
\label{tau03}\\
\tau^\pm_{(+1)}&=&
\vev{\Psi_0\cdot\prod_pW_{u_p}(\mu_p)
\prod_j\({ 1 \pm y_j W_2(\mu_j)}\)\Psi_0^{-1}\cdot p_{-1} }
\label{tau+13}\\
\tau^\pm_{(-1)}&=&
\vev{p_1\cdot \Psi_0\cdot\prod_pW_{u_p}(\mu_p)\prod_j
\({ 1 \pm y_j W_2(\mu_j)}\)\cdot \Psi_0^{-1}}
\label{tau-13}
\fin
with $\Psi_0$ the vacuum wave function,
\debut
\Psi_0= \({\frac{B}{\rho}}\)^{E_+} B^{E_-}= \({\frac{\rho}{A}}\)^{E_-}A^{-E_+}
\non
\fin
We recall that $A,~B$ are defined by $A(z_+)=a(z_+)+1/2$ and
$B(z_-)=b(z_-)+1/2$. They are related to $\rho$ by
$(A + B-1)=\rho$.
 
To compute $\tau^\pm$ one therefore needs to know how to conjugate
vertex operators $W_u(\mu)$ with $E_\pm$.  One has:
\debut
\Psi_0\cdot W_u(\mu)\cdot \Psi_0^{-1} = \rho^\frac{u^2}{4}
\({ \frac{\mu_+(A)}{\mu_-(B)} }\)^\frac{u^2}{2}\cdot
 W_u\({\frac{\mu_+(A)\mu_-(B)}{\mu} }\)\cdot \label{wmuAB}
\fin
where
\debut
\mu_+(A)^{2}= \frac{\mu^2}{\mu^2 + (1-\mu^2)A} \quad;\quad
\mu_-(B)^{-2}= \frac{\mu^{-2}}{\mu^{-2} + (1-\mu^{-2})B}
\label{muAmuB}
\fin
These formula are proved in Appendix D.
They are derived using the fact that $V_u(\mu)=\mu^{-\frac{u^2}{4}} W_u(\mu)$
is a primary field of conformal weight $(\frac{u^2}{4})$.
It implies that under a diffeomorphism in $\mu$, the fields $V(\mu)$
transform as $(\frac{u^2}{4})$-forms.
The functions $\mu_+(A)$ and $\mu_-(B)$ are
simply the integral curves of the vector fields $E_\pm$.
Notice that the vertex operators at $\mu=1$ satisfy simple
relations:
\debut
\Psi_0\cdot W_u(1)\cdot \Psi_0^{-1}
= \rho^{\frac{u^2}{2}}~ W_u(1) \label{w1}
\fin
 
Using eqs.(\ref{wmuAB})
and the reordering formula (\ref{reorder}),
the final results we get for the $\tau$-functions are:
\debut
\tau^\pm_0(z_+,z_-) &=&\Om_{\{u_p\}}(z_+,z_-)\cdot
\vev{\prod_pW_{u_p}\({\mu_p(z_+,z_-) }\)\prod_j\Bigl({ 1 \pm Y_j(z_+,z_-)
W_2\({ \mu_j(z_+,z_-)}\) }\Bigr)  } \label{ouf}\\
\tau^\pm_{(+1)}(z_+,z_-)&=&\Om_{\{u_p\}}(z_+,z_-)\cdot
\vev{\prod_pW_{u_p}\({\mu_p(z_+,z_-) }\)\prod_j\Bigl({ 1 \pm Y_j(z_+,z_-)
W_2\({ \mu_j(z_+,z_-) }\) }\Bigr) \cdot p_{-1} } \non\\
\tau^\pm_{(-1)}(z_+,z_-)&=&\Om_{\{u_p\}}(z_+,z_-)\cdot
\vev{p_1\cdot \prod_pW_{u_p}\({\mu_p(z_+,z_-) }\)\prod_j\Bigl({ 1 \pm Y_j(z_+,z_-)
W_2\({\mu_j(z_+,z_-)}\) }\Bigr) } \non
\fin
with
\debut
\Om_{\{u_p\}}(z_+,z_-)&=& \prod_p \({\rho^\half\cdot
\frac{\mu_{p,+}(A)}{\mu_{p,-}(B)}}\)^\frac{u_p^2}{2},\non\\
\mu(z_+,z_-)&=& \frac{\mu_{+}(A)\mu_{-}(B)}{\mu}\label{oufouf}\\
Y_j(z_+,z_-) &=& y_j\cdot \rho\cdot
\({\frac{\mu_{j,+}(A)}{\mu_{j,-}(B)}}\)^2\non
\label{final}
\fin
They  can be easily evaluated using the Wick's theorem (\ref{wick}) and the
commutation relations (\ref{p1w}). Cf. Appendix D.
They admit simple expressions as determinants.
We recall that the expressions for $\hat \sig,~\varphi$ and $X_{(\pm1)}$
are extracted from eqs.(\ref{ouf}) using eqs.(\ref{tau2},\ref{tauy1su2},
\ref{tauy2su2}).

\subsection{Vertex operators and the dual fields.}
Vertex operators can also be used to compute algebraically
the dual variables. The point is that in the representation
$\La_\pm$ the element $\hat \CE^*$ entering
the magic formula (\ref{magic3}) admits a simple expression
in terms of vertex operators. Namely, in the representation
$\La_\pm$ we have:
\debut
\hat \CE^*= \hat \CE^*_+-\hat \CE^*_- = \pm \frac{i}{2}~ W_2(1) \non
\fin
Thus we may express the dual fields as:
\debut
N^* \pm \frac{i}{\De^*} = \pm i
\frac{\bra{\La_\pm}\Psi_0\cdot(g_-^{-1}h^{-1}W_2(1)hg_+)
\cdot\Psi_0^{-1}\ket{\La_\pm}}{
\bra{\La_\pm}\Psi_0\cdot(g_-^{-1}g_+)\cdot\Psi_0^{-1}\ket{\La_\pm}}
\label{x*w2}
\fin
The conformal factor $\hat \sig^*$ is given by eq.(\ref{dualtau}).
 
In eq.(\ref{x*w2}) we have explicitely reintroduced the parameter $h\in SO(2)$
corresponding to the $SO(2)$ gauge transformation $g_\pm \to hg_\pm$.
Recall that eq.(\ref{magic3}) and therefore eq.(\ref{x*w2}) is valid only
in the triangular gauge. Thus eq.(\ref{x*w2}) is only valid for
a specific valid of $h$.
 
We now explain an algebraic way to fixe the $SO(2)$ parameter $h$
in order to ensure the triangular gauge (\ref{QPtri}).
This will be done by comparing the expression of the connexion
$Q_\pm+P_\pm$ in the original and dual variables.
First imagine computing (\ref{x*w2}) for any $h\in SO(2)$.
This gives two functions $(N^*_h,\De^*_h)$ depending on $h$.
Define then $(Q_\pm+P_\pm)_h$ by:
\debut
(Q_\pm+P_\pm)_h = -\half (\rho\De^*_h)^{-1}\d_\pm(\rho\De^*_h) \sig^z
\pm (\De^*_h\d_\pm N^*_h)\cdot \sig^-
\label{QPh}
\fin
Of course $(Q_\pm+P_\pm)_h$ will be solution of Einstein's equations only
for $h$ equals to the specific value corresponding to the triangular gauge.
Compute now the connexion $Q_\pm+P_\pm$ in the triangular gauge in
the original variables $(\De,N)$ using eq.(\ref{drvac}).
Its expression is:
\debut
Q_\pm+P_\pm = \frac{d_\pm}{2}
\({X^1_{(\mp 1)} \cos\varphi_\mp - X^2_{(\mp 1)} \sin\varphi_\mp }\)\cdot\sig^z
+ d_\pm\({X^1_{(\mp 1)} \sin\varphi_\mp + X^2_{(\mp 1)}
\cos\varphi_\mp }\)\cdot\sig^-\label{qdrtri}
\fin
Imposing the triangular gauge demands choosing $\varphi_\pm$ with
$\varphi=\varphi_+-\varphi_-$ such that:
\debut
d_\pm\({X^1_{(\mp 1)} \sin\varphi_\mp + X^2_{(\mp 1)} \cos\varphi_\mp }\)
= - \d_\pm \varphi_\pm \non
\fin
We will not need to solve this equation explicitly.
Now the parameter $h$ corresponding to the triangular gauge
is by definition such that $(Q_\pm+P_\pm)_h=Q_\pm+P_\pm$.
This yields two equations determining $\varphi_\pm$:
\debut
\frac{X^1_{(\mp 1)} \cos\varphi_\mp - X^2_{(\mp 1)} \sin\varphi_\mp}{
X^1_{(\mp 1)} \sin\varphi_\mp + X^2_{(\mp 1)} \cos\varphi_\mp}
= \mp \frac{(\rho\De^*_h)^{-1}\d_\pm(\rho\De_h^*)}{
\De^*_h\d_\pm N^*_h}
\equiv \mp \frac{A_\pm^h}{B_\pm^h}
\label{hphi}
\fin
It allows us to expresse algebraically $e^{i\varphi_\pm}$
in terms of $A_\pm^h=(\rho\De^*_h)^{-1}\d_\pm(\rho\De^*_h)$,
$B_\pm^h=\De^*_h\d_\pm N^*_h$ and of the known $\tau$ functions.
Recalling that $e^{i(\varphi_+-\varphi_-)/2}=\tau^-_0/\tau^+_0$
gives the desired constraint on $h$:
\debut
\frac{(A_-^h-iB_-^h)(A_+^h-iB_+^h)}{(A_-^h+iB_-^h)(A_+^h+iB_+^h)}
= \({\frac{\tau^-_0}{\tau_0^+}}\)^4
\({ \frac{\tau^-_{(+1)}\tau^+_{(-1)}}{\tau^+_{(+1)}\tau^-_{(-1)}} }\)
\label{hconst}
\fin
This algebraic procedure fixing $h$ to ensure the triangular
gauge is a little heavy and could probably be simplified.
 
We explain in Appendix E how for elements $g_\pm$ of the form given in
eqs.(\ref{wuh}) or (\ref{gsol}), the product $(g_-^{-1}h^{-1}W_2(1)hg_+)$
can again be written as product of vertex operators for any $h\in SO(2)$.
Once this is done the expectation values (\ref{x*w2}) can
be computed using eq.(\ref{wmuAB}) and the Wick theorem.
\bigskip
 
{\bf To summarize:} Vertex operators which are operators acting on an
auxiliary Fock space provide a direct way to compute the $\tau$
functions.
 
\section{Examples of explicit formulas for the metric.}
Here we describe the simplest possible applications of the
algebraic methods we have developped. Most of these simple
examples of solutions have already been found eg. in ref.\cite{zaka}.

\subsection{Formula using the variables $N,~\De$.}
We recall that in the variables $N,~\De$ the metric is
given by the formula
\debut
ds^2=2\rho^\half e^{2\hat \sig} dAdB + \rho\De^{-1} dx^2
+\rho\De(dy - Ndx)^2 \label{again1}
\fin
where the connexion $(Q_\pm+P_\pm)$ which is given in terms of $(N,~\De)$ by
$Q_\pm+P_\pm =\half (\De^{-1}\d_\pm\De)\cdot \sig^z
+ \De(\d_\pm N)\cdot \sig^-$ is evaluated using eqs.(\ref{drvac}).
 
Let us consider the simple cases where the metric is diagonal, ie. $N=0$.
This corresponds to choose to dress by elements of the form:
\debut
g_-^{-1} g_+ = \prod_p W_{u_p}(\mu_p) \label{gdiag}
\fin
The $\tau$-functions are then:
\debut
\tau^\pm_0 &=& \Om_{\{u_p\}}(z_+,z_-)\cdot
\prod_{p<q} \({\frac{\mu_p(z_+,z_-)-\mu_q(z_+,z_-)}{
\mu_p(z_+,z_-)+\mu_q(z_+,z_-)}}\)^{u_pu_q/2} \non\\
\tau_{(+1)}^\pm &=& \tau^\pm_0\cdot \sum_p u_p \mu_p^{-1}(z_+,z_-) \non\\
\tau_{(-1)}^\pm &=& -\tau^\pm_0\cdot \sum_p u_p \mu_p^{1}(z_+,z_-) \non
\fin
Therefore by definitions (\ref{tau2},\ref{tauy1su2},\ref{tauy2su2}),
$\varphi=0$, $X^2_{(\pm1)}=0$ and $X^1_{(\pm 1)} = \sum_p u_p
\mu_p^{\mp1}(z_+,z_-)$. Eqs. (\ref{drvac}) then give us the
formula for the connexion $(Q_\pm+P_\pm)$:
\debut
Q_\pm+P_\pm = \frac{d_\pm}{2}\({ \sum_p u_p\mu_p^{\pm1}(z_+,z_-)}\)
\sig^z \non
\fin
By integration it corresponds to
\debut
N=0\quad,\quad
\De= {\rm const.}~ \prod_p\({\frac{1-\mu_p(z_+,z_-)}{1+\mu_p(z_+,z_-)}}\)^{u_p}
\label{delta1}
\fin
The conformal factor $\hat \sig$ is computed using eq.(\ref{tautau}):
\debut
e^{2\hat\sig}=\tau_0^+\tau_0^-= \prod_p \({\rho^\half\cdot
\frac{\mu_{p,+}(A)}{\mu_{p,-}(B)}}\)^{u_p^2}\cdot
\prod_{p<q} \({\frac{\mu_p(z_+,z_-)-\mu_q(z_+,z_-)}{
\mu_p(z_+,z_-)+\mu_q(z_+,z_-)}}\)^{u_pu_q}    \label{sig1}
\fin
Inserting (\ref{delta1},\ref{sig1}) in (\ref{again1})
gives simple solutions of Einstein's equations.

\subsection{Formula using the dual variables $N^*,~\De^*$.}
In the dual variables $N^*,~\De^*$ the metric is
given by the formula
\debut
ds^2=2\rho^\half e^{2\hat \sig^*} dAdB + \rho\De^{*~-1} dx^2
+\rho\De^*(dy - N^*dx)^2 \label{again2}
\fin
We recall that in the vertex operator representations
the dual potentials which are determined using eq.(\ref{x*w2}).
The conformal factor $\hat \sig^*$ is given by eq.(\ref{dualtau}).
\bigskip

We first consider the diagonal cases, $N^*=0$,
which for example correspond to dress by the elements:
\debut
 g_\pm = \prod_p \exp\({ u_p\frac{\sig^z}{2}\otimes 
\log\({ \frac{1+ (t/\mu_p)^{\pm1}}{1-(t/\mu_p)^{\pm1}} }\) }\) 
\label{gab}
\fin
The product $g_-^{-1}g_+$ is equal up to a scalar factor
to the product of vertex operators (\ref{gdiag}).
In eq.(\ref{gab}), we fixe the gauge freedom associated
to the $g_\pm \to h g_\pm$, $h\in SO(2)$  gauge transformation.
It may be check that this corresponds to the 
triangular gauge for the connexion $(Q_\pm+P_\pm)$.
Commuting the elements $g_\pm$ through $W_2(1)$
using the standard $e^xe^y=e^{[x,y]}e^ye^x$ formula,
one finds that
\debut
g_-^{-1} W_2(1) g_+ = \prod_p W_{u_p}(\mu_p)\cdot W_2(1) \non
\fin
up to an irrelevent scalar factor. 
Therefore, using eqs.(\ref{x*w2}) and (\ref{wmuAB}) one obtains:
\debut
N^* \pm \frac{i}{\De^*} = \pm i \rho~
\frac{\vev{\prod_p W_{u_p}(\mu_p(z_+,z_-))\cdot W_2(1)}}{
\vev{\prod_p W_{u_p}(\mu_p(z_+,z_-))}}
\non
\fin
which gives
\debut
N^*=0\quad,\quad
\rho\De=\inv{\De^*}= {\rm const}\cdot\rho
\prod_p\({\frac{1-\mu_p(z_+,z_-)}{1+\mu_p(z_+,z_-)}}\)^{u_p} \non
\fin
Of course this corresponds to the metrics we find using the 
original variables $N,~\De$.
\bigskip

The non diagonal cases, ie. $N\not=0$ or $N^*\not =0$, correspond for
example to elements of the form:
\debut
g_-^{-1}g_+= g_-^{-1}(1)\cdots g_-^{-1}(M)\cdot g_+(M)\cdots g_+(1)
\non
\fin
where $g_\pm(j)$ are given in eq.(\ref{defg-g+}). 
As we have explained the computations in the non diagonal cases
are a little more involved since they require fixing the triangular gauge
(\ref{QPtri}). In the original variables $(\De,~N)$ this requires
choosing $\varphi_\pm$ appropriately. 
In the dual variables $(\De^*,~N^*)$, this requires choosing the gauge
$g_\pm \to hg_\pm$ in order to apply formula (\ref{magic3}),
or alternatively eq.(\ref{x*w2}). An algebraic procedure 
to solve these gauge fixing problems was given in Section 6.3.
\bigskip

We leave to the reader the pleasure to check that these formulas
give local solutions of Einstein's equations.
Not all these solutions correspond to physically acceptable solutions:
one has to eliminate naked singularities, 
their global properties have to be analysed, etc ...

\bigskip
\vskip 1.0 truecm

{\bf Acknowledgements:} This is research is supported in part
by the European TMR contract ERBFMRXCT960012.
We thank Olivier Babelon for his interest for this project.

\section{Appendix A. Coupling to Liouville and 
another use of the semi-direct product $Vir\times \CH_{taf}$. }
Here we give another example of the use of the algebra
$Vir\times \CH_{taf}$ for constructing Lax pairs. 
It consists in coupling sigma models to Liouville theory.
The Liouville theory will be linked to the Virasoro algebra whereas
the sigma models will be linked to the affine algebra.
The semi direct product structure provides the way to couple
these two models.

Namely, let us define another Lax connexion $A_\pm'$ by
\debut
A'_\pm = \mp (\d_\pm \Phi) L_0 + e^{-\Phi} L_{\pm 1}
+ Q_\pm + P_\pm\otimes t^{\pm 1}
\mp (\d_\pm \sig)\frac{k}{2} \non
\fin
It defines a compatible Lax connexion. Its
zero curvature condition is then equivalent to the equations:
\debut
\d_+\d_-\Phi- e^{-2\Phi} &=& 0 \non\\
\d_+Q_--\d_-Q_+ +[Q_+,Q_-]+ [P_+,P_-] &=& 0 \non\\
D_+\({e^{\Phi/2}P_-}\) + \half e^{-\Phi/2}P_+ &=&0 \non\\
D_-\({e^{\Phi/2}P_+}\) + \half e^{-\Phi/2}P_- &=&0 \non
\fin
The equation for $\Phi$ is the Liouville equation.

An analysis similar to the one we have described for the reduced gravity
can be done. In particular dressing transformations can be introduced.
The dressing group is actually bigger than for the reduced gravity since
it is isomorphic to the Borel subalgebras of $Vir\times \CH_{taf}$.
One may imagine other applications to integrable models of the
semi direct product structure.

\section{Appendix B. ``Abelian solutions" and their wave functions.}
Here we study in detail a very simple class of solutions.
They provide a simple exercise in which one can test the
various constructions.

When $P_\pm=0$, solutions are trivial since then the equations
reduce to $\d_-Q_+-\d_+Q_-+[Q_-,Q_+]=0$. Therefore, $(Q_\pm+P_\pm)\in {\bf h}$
is a pure gauge which can be gauged away.
More interesting solutions need $P_\pm\not= 0$.
For any $\bu \in \bf r$, a simple solution of the equations of motion is given by:
\debut   
Q_\pm^\bu&=&0 \non\\
P^\bu_\pm&=& d_\pm \bu = \rho^{-1}\d_\pm\rho~ \bu,\quad \bu\in{\bf r}
\label{vac}\\
{\hat\sig}^\bu &=& \frac{\bu^2}{2} \log\rho \non
\fin
with $\bu^2=tr(\bu\bu)$. Note that $\d_\pm{\hat\sig}^\bu= \frac{\bu^2}{2}d_\pm$.
We call these solutions, abelian solutions.
The Lax connexions for these abelian solutions are:
\debut   
A^\bu_{\pm}= d_\pm\({ \pm E_\pm + \bu\otimes t^{\pm1}\mp \frac{\bu^2}{4}k }\)
\label{laxvac}
\fin
Their wave functions $\Psi_\bu$ are given by:
\debut   
\Psi_\bu(z_+,z_-) &=& h^\bu_+~\cdot
\({\frac{b(z_-)+\inv{2}}{\rho}}\)^{E_+}~ \({b(z_-)+\inv{2}}\)^{E_-}~
\cdot \rho^{\frac{\bu^2}{4}k} ~\cdot {h_+^\bu}^{-1}
\label{psivac1}\\
&=& h^\bu_-~\cdot
\({\frac{\rho}{a(z_+)+\inv{2}}}\)^{E_-}~ \({a(z_+)+\inv{2}}\)^{E_+}~
\cdot \rho^{-\frac{\bu^2}{4}k} \cdot {h_-^\bu}^{-1}
\label{psivac2}
\fin
with
\debut
h^\bu_\pm = \exp\({ \bu\otimes \log\({\frac{1+t^{\pm1}}{1-t^{\pm1}} }\) }\)
\quad \in \exp(\CB_\pm) \label{hvac}
\fin
Notice that $h^\bu_\pm$ belong to opposite Borel subgroups.
We choose to normalize $\Psi_\bu$ such that it reduces to the
identity at the point where $a(z_+)=b(z_-)=1/2$.
One may compute the fields $\xi_\la$ and $\bar \xi_\la$ for the vacuum
solutions and effectively check that they are chiral.
\bigskip

Let us give the proof of the formula (\ref{psivac1},\ref{psivac2}).
Let us decompose $\Psi_\bu$ as $\Psi_\bu= \hat \psi~ \psi_V$.
Then $\hat \psi$ satisfies eq.(\ref{akmbis}) with ${\hat A^{km}}$ given
by the abelian Lax connexion, ie.
\debut
\d_\pm \hat \psi \pm [E_\pm,\hat \psi]+
d_\pm\({\bu\otimes t^{\pm1} \mp \frac{\bu^2}{4}k }\) \hat \psi=0
\label{Blin}
\fin
To derive eq.(\ref{psivac2}) decompose $\hat \psi$ as
$\hat \psi= h_- h_0 h_+ e^{\eta k}$ where $h_\pm$ belong to the two
Borel subgroups and $h_0\in H$. This is just a Gauss decomposition.
Then eq.(\ref{Blin}) can be rewritten as:
\debut
h_-^{-1}\cdot\d_\pm h_- + \d_\pm(h_0 h_+)\cdot(h_0 h_+)^{-1} 
+ \d_\pm\eta\cdot k &=& \mp d_\pm \({ h_-^{-1}E_\pm h_- - 
(h_0 h_+)E_\pm (h_0 h_+)^{-1} }\) \non\\
&~&~~- d_\pm~h_-^{-1}\(\bu\otimes t^{\pm1} \mp \frac{\bu^2}{4}k \)h_- \non
\fin
This equation can be analysed degree by degree since by definition
$h_-^{-1}\cdot\d_\pm h_-$ is of negative degree, 
$\d_\pm h_+\cdot h_+^{-1}$ is of positive degree and
$\d_\pm h_0 \cdot h_0 ^{-1}$ of degree zero.

The abelian solutions in eq.(\ref{psivac2}) is
such that $(h_0h_+)=1$ and $h_-^{-1}\cdot\d_\pm h_-=0$ and 
$\d_\pm\eta = - d_\pm \frac{\bu^2}{4}$.
The specific form of $h_-$ given in eq.(\ref{hvac}) is such that:
\debut
h_-E_- h_-^{-1} - E_-  &=& -\bu \otimes t^{-1} \non\\
h_-E_+ h_-^{-1} - E_+  &=& \bu \otimes t - \bu^2\cdot 
\frac{k}{2} \non
\fin
This proves eq.(\ref{psivac2}). The alternative formula
(\ref{psivac1}) is proved similarly but doing the Gauss decomposition
in the other order.
\bigskip

Let us now show that all the abelian solutions 
for different values of $\bu$ are related
by dressing transformations, ie. they belong to the same orbit of the
dressing group. More precisely, let $\Psi_0$ be the vacuum wave
function corresponding to $\bu=0$, cf. eq.(\ref{psizero}).
Then, dressing $\Psi_0$ by $(h^\bu_-{}^{-1}h^\bu_+)$ gives the
wave function $\Psi_\bu$:
\debut   
D_{(h^\bu_-,h^\bu_+)}(\Psi_0)&\equiv &(\Psi_0h^\bu_-{}^{-1}h^\bu_+\Psi_0^{-1})_\pm
\cdot \Psi_0\cdot  h_\pm^\bu{}^{-1}\non\\
&=&\rho^{\pm\bu^2k/4}\cdot h_\pm^\bu\cdot \Psi_0\cdot h_\pm^\bu{}^{-1}
= \Psi_\bu \label{drvac0}
\fin
This is equivalent to the statement that:
$(\Psi_0h^\bu_-{}^{-1}h^\bu_+\Psi_0^{-1})= h^\bu_-{}^{-1}h^\bu_+\cdot
\rho^{\bu^2k/2}$ namely that both expressions for $\Psi_\bu$ are equal. 
One may also verify that the
formulas (\ref{dresq},\ref{dresp}) for the dressed connexion
reproduce the formula (\ref{vac}).
\bigskip

\section{Appendix C. Dressing by $E_\pm$.}
Here we describe how to extend the group of dressing transformations
with the elements $E_\pm$, ie. we describe the dressing
by elements of the form $e^{a_-E_-}e^{a_+E_+}$.
The point about these transformations is that they modify
the factor $\rho$. 

Since we have  already described dressing transformations in detail 
we will be more schematic. 
We need to extend a little the factorization problem (\ref{factor2}).
It is defined as before for the component in the affine group but for
the component along $E_\pm$ we choose to put $E_-$ into $g_-$ and
$E_+$ into $g_+$. Ie. we are still decomposing with respect
to the Borel subalgebras. Thus consider $g=g_-^{-1}g_+=e^{a_-E_-}e^{a_+E_+}$.
Dressing transformations are again defined on the wave function by:
\debut
D(\Psi) = \hat \Th_\pm~\Psi~ g_\pm^{-1} \label{extdres}
\fin
where $\hat \Th_\pm$ is determined by factorizing $(\Psi g\Psi^{-1})$:
\debut
\hat \Th_\pm = (\Psi e^{a_-E_-}e^{a_+E_+}\Psi^{-1})_\pm
\equiv e^{\pm \chi_\pm E_\pm}~ \Th_\pm \label{extth}
\fin
with $\Th_\pm\in \exp(\CB_\pm\oplus \Cmath k)$.
Eq.(\ref{extth}) serves as definition of $\chi_\pm$.
It should be noted that the fact that $ \Ga = \hat \Th_-^{-1}\hat \Th_+$
is solution of $\d\Ga + [A,\Ga]=0$ implies that $\chi_-$ and $\chi_+$
differ by a constant:
\debut
\chi_- + {\rm const.} = \chi_+ + {\rm const.} \equiv \chi \non
\fin

It is easy to check that the form of the Lax connexion is 
preserved by the transformation (\ref{extdres}).
Thus this transformation is a symmetry of the equations of motion.
By analysing the way the Lax connexion transforms one finds that
$\rho$ transforms as:
\debut
\rho \to e^{-\chi} \rho \label{extrho}
\fin
The action on the other fields induced by these transformations
is also non trivial.

\section{Appendix D. Evolution rules for vertex operators.}
Here we give the proof of the formula (\ref{wmuAB}).
They are derived using the fact that 
for any $u$ the operators $\mu^{-\frac{u^2}{4}} W_u(\mu)$ 
are primary fields of weight $(\frac{u^2}{4})$.

More precisely, recall that the Virasoro generators $L_n$ 
in the level one representations $\La_\pm$ are given by:
\debut
\sum_n \({ L_n - \inv{16}\de_{n,0} }\)~ \mu^{-2n-2}
= \inv{4} :(i\d_\mu Z)^2: \non
\fin
Their commutation relations with the vertex operators $W_u(\mu)$ are:
\debut
\[ L_n, W_u(\mu) \] = \frac{\mu^{2n}}{2} \({ 
\frac{u^2}{2} n + \mu\d_\mu }\) W_u(\mu) \label{Acom1}
\fin
In particular,
\debut
\[ E_\pm , W_u(\mu) \] = \half \({ \mp \frac{u^2}{2} \mu^{\pm 2}
+ \mu(1-\mu^{\pm 2})\d_\mu }\) W_u(\mu) \label{Acom2}
\fin
These commutation relations are not the usual commutation relations for
primary fields.  To recover the primary fields, one has to change variable
from $\mu $ to $z=\mu^2$. Then, the primary fields $\Phi_u(z)$ are
defined by $\Phi_u(z)= \mu^{-u^2/2} W_u(\mu)$. They satisfy:
\debut
\[ L_n,\Phi_u(z)\] = z^n\({\frac{u^2}{4}(n+1)+z\d_z}\) \Phi_u(z) \non
\fin
This implies that under a diffeomorphism $z\to f(z)$, the fields
$\Phi_u(z)$ transform as:
\debut
\Phi_u(z) \to (\d_zf(z))^\frac{u^2}{4}~ \Phi_u(f(z)) \non
\fin
To the diffeomorphism $z\to f(z)$ corresponds a diffeomorphism
$\mu \to F(\mu)=\sqrt{f(\mu^2)}$, and the vertex operators $W_u(\mu)$ transform as:
\debut
\mu^{-\frac{u^2}{4}} W_u(\mu) \to~
(\d_\mu F(\mu))^{\frac{u^2}{4}}~(F(\mu))^{-\frac{u^2}{4}}~
W_u(F(\mu)) \label{Atrans}
\fin

It is now easy to find the formula (\ref{conjA},\ref{conjB}).
As it can be seen from eq.(\ref{Acom2}),
the transformations generated by $E_\pm$ correspond to
diffeomorphisms is $\mu$ generated by the vector fields
$E_\pm(\mu)= \frac{1}{2}(1-\mu^{\pm2})\mu\d_\mu$.
The two functions $\mu_+(A)$ and $\mu_-(B)$ which specify the
diffeormophism associated to the conjugation formula (\ref{conjA},\ref{conjB})
are simply the integral curves of the vector fields $E_\pm(\mu)$.
Hence, $\mu_+(A)$ is defined by the differential equation
\debut
A\d_A\mu_+(A)=-\half(\mu_+(A)-\mu_+(A)^3) \label{Aflow}
\fin
with the initial condition $\mu_+(A=1)=\mu$.
Similarly, $\mu_-(B)$ is defined by the differential equation,
\debut
B\d_B \mu_-(B)= \half(\mu_-(B)-\mu_-(B)^{-1}) \label{Bflow}
\fin
with the initial condition $\mu_+(B=1)=\mu$.
The solutions of these equations are those given in eq.(\ref{muAmuB}).
Applying the transformation rules (\ref{Atrans}) with
$F(\mu)$ equals either $\mu_+(A)$ or $\mu_-(B)$ gives 
teh following conjugation formula:
\debut
A^{-E_+}\cdot W_u(\mu)\cdot A^{E_+} &=&
A^{\frac{u^2}{4}} \({\frac{\mu_+(A)}{\mu}}\)^{\frac{u^2}{2}} W_u(\mu_+(A)) \label{conjA}\\
B^{E_-}\cdot W_u(\mu)\cdot B^{-E_-} &=&
B^\frac{u^2}{4} \({\frac{\mu}{\mu_-(B)}}\)^\frac{u^2}{2} W_u(\mu_-(B)) \label{conjB}
\fin
In particular, eqs.(\ref{conjA},\ref{conjB}) and the reordering formula
(\ref{reorder}) imply eq.(\ref{wmuAB}) used in the text.

Finally we quote the commutation relations bewteen $p_n$
and  the vertex operators:
\debut
\[ p_n, W_u(\mu)\]= - u~\mu^n~W_u(\mu)
\label{p1w}
\fin
They are used to compute the $\tau$-functions.
Besides eq.(\ref{wick}), a sample of formula needed to compute the
$\tau$-functions is:
\debut
\vev{\prod_{j=1}^M(1\pm y_j W_2(\mu_j))}= \sum_{p=0}^M (\pm)^p
\sum_{k_1<\cdots<k_p} y_{k_1}\cdots y_{k_p}
\prod_{k_i<k_j} \({\frac{\mu_{k_i}-\mu_{k_j}}{\mu_{k_i}+\mu_{k_j}}}\)^2
\non\\
\vev{\prod_j W_{u_j}(\mu_j)\cdot p_{-1}}
= \prod_{i<j} \({\frac{\mu_{i}-\mu_{j}}{\mu_{i}+\mu_{j}}}\)^{u_i\cdot u_j/2}
\cdot \sum_j u_j~ \mu_j^{-1} \non\\
\vev{p_1\cdot\prod_j W_{u_j}(\mu_j)\cdot} 
= \prod_{i<j} \({\frac{\mu_{i}-\mu_{j}}{\mu_{i}+\mu_{j}}}\)^{u_i\cdot u_j/2} 
\cdot \sum_j u_j~ \mu_j \label{listop}
\fin

\section{Appendix E. Relation between $\hat \CE^*_\pm(\mu)$ and vertex operators.}
Here we explain the relation (\ref{g-g+w}) using result from
\cite{babe2} and we explain how to compute
$(g_-^{-1}h^{-1} W_2(1)hg_+)$ for any $h\in SO(2)$.
Let us recall the relation (\ref{g-g+w}):
\debut
g_-^{-1}g_+= g_-^{-1}(1)\cdots g_-^{-1}(M)\cdot g_+(M)\cdots g_+(1)=
\prod_{j=1}^M\({ 1 \pm y_j W_2(\mu_j)}\) \label{g-g+wbis}
\fin
where $g_\pm(j)$ are the elements defined in eq.(\ref{defg-g+}).
Recall that eq.(\ref{g-g+wbis}) is an equation valid in the
representation $\La_\pm$.
 
The result of \cite{babe2} can be summarized as follows:
The parameters $(s_j,r_j,v_j)$ entering in the definition
of $g_\pm(j)$ are recursively computed in terms of
the parameters $(y_j,\mu_j)$. First $(s_j,r_j)$ are
determined as a functions of the $(y_p,\mu_p)$ with $p\leq j$ by,
\debut
\exp\({\sum_{j=1}^p \({r_j\pm i\frac{s_j}{2}}\) }\)
= \vev{\prod_{j=1}^p(1 \pm y_j W_2(\mu_j))} \label{bb01}
\fin
Then, writing $s_j=-\eta_j+\eta_{j-1}$ with $\eta_0=0$,
the parameters $v_j$ are equals to $v_j=-\rho_j+\rho_{j-1}$
with $\rho_0=0$ and
\debut
\rho_j(y_p)=\eta_j(\beta_{j+1;p} y_p) \non
\fin
where $\beta_{n;p}=\({\frac{\mu_n-\mu_p}{\mu_n+\mu_p}}\)$.
Notice that $(s_j,r_j)$ depend on $(y_p,\mu_p)$ with $p\leq j$
only but that $v_j$ depends on $(y_p,\mu_p)$ with $p\leq j$
but also on $\mu_{j+1}$.
 
The parameter $v_M$ is irrelevent in eq.(\ref{g-g+wbis}) since
it cancels in the product $g_-^{-1}g_+$. Changing
its value corresponds to the gauge transformation
$g_\pm \to h g_\pm$ with $h\in SO(2)$.
 
The elements $g_\pm(j)$ satisfy nice commutation relations
with the vertex operators which are needed to
algebraically evaluate the dual fields. Let us introduce
the line vector $E(\mu_0)$ with operator entries:
\debut
E(\mu_0)= \({1, \pm\half W_2(\mu_0), \pm\half W_2(-\mu_0),
i\mu_0\frac{d Z(\mu_0)}{d\mu_0} }\)
\non
\fin
where $W_2(\mu_0)$ and $i\mu_0\frac{d Z(\mu_0)}{d\mu_0}$ are the vertex operators
used in the representation (\ref{repv}) of the affine algebra.
The sign $\pm$ depends whether we consider the representation $\La_+$
or $\La_-$. It transforms simply under $SO(2)$ conjugation:
\debut
h^{-1} \cdot E(\mu_0) \cdot h = E(\mu_0)\cdot V_h,
\quad h\in SO(2)
\label{Eso2}
\fin
where $V_h$ is an explicitely known $\Cmath$-number four dimensional vector
depending only on $h$.
Moreover, as shown in \cite{babe2}, $E(\mu_0)$ also transforms
in a simple way by conjugation with the $g_+(j)$:
\debut
 g_+(j)^{-1}\cdot E(\mu_0) \cdot g_+(j) = E(\mu_0)\cdot R_0(j)
\label{roj}
\fin
with $R_0(j)$ is an explicitely known $\Cmath$-number $4\times 4$ matrices
functions of $v_j,r_j$ and $\mu_0,\mu_j$.
 
Formula (\ref{Eso2},\ref{roj}) gives a way to compute
the product $g_-^{-1}h^{-1} W_2(\mu_0) h g_+$ for $h\in SO(2)$
and $g_\pm= g_\pm(M)\cdots g_\pm(1)$ as follows:
\debut
g_-^{-1}h^{-1} W_2(\mu_0) h g_+ &=&
g_-^{-1}\cdot (E(\mu_0)\cdot V_h)\cdot g_+ \non\\
&=& g_-^{-1}g_+\cdot (E(\mu_0)\cdot R_0(1)R_0(2)\cdots R_0(M)V_h) \label{bab}
\fin
Furthermore, $g_-^{-1}g_+$ may now be replaced by its vertex operator
representation (\ref{g-g+wbis}). For $\mu_0=1$ this is the formula
which allows to compute the dual fields only terms of vertex operators.


\begin{thebibliography}{}
%
%
\bibitem{ger} R. Geroch, J. Math. Phys. 13 (1972) 394.

\bibitem{zaka} V.A. Belinskii and V.E. Zakharov, Sov. Phys. JETP 48 (1978) 985.

\bibitem{mai1} D. Maison, Phys. Rev. Lett.  41 (1978) 521.
 
\bibitem{julia} B. Julia, {\it Infinite dimensional algebras in physics}, in
"John Hopkins workshop on current problems in particle physics: Unified theories
and beyond", Johns Hopkins University, Baltimore (1981).

\bibitem{mais} P. Breitenlohner and D. Maison, Ann. Inst. H. Poincare, 46 (1987) 215.

\bibitem{giha} G. Gibbons and S. Hawking, Com. Math. Phys. 66 (1979) 291.

\bibitem{creju} E. Cremmer and B. Julia, Nucl. Phys. B1159 (1979) 141.

\bibitem{julanico} B. Julia and H. Nicolai, hep-th/9608082, Nucl. Phys. 
{\bf B482} (1996) 431.

\bibitem{zakasha} V.E. Zakharov and A.B. Shabat, Funct. Anal. Appl. 13 (1979) 166

\bibitem{semen} M. Semenov-Tian-Shansky, Publ. RIMS 21 (1985) 1237.

\bibitem{kinn} W. Kinnersley, J. Math. Phys. 18 (1977) 1529. 

\bibitem{mato} W. Woodhouse and L. Mason, Nonlinearity 1 (1988) 73. 

\bibitem{nico} cf. eg H. Nicolai, D. Korotkin and H. Samtleben, 
{\it Integrable classical and quantum gravity}, hep-th/9612065,
and references therein.

\bibitem{lewi} J. Lepowsky and R. Wilson, Commun. Math. Phys. 62 (1978) 43.

\bibitem{koni} D. Korotkin and H. Nicolai, Phys. Rev. Lett. 74 (11995) 1272.

\bibitem{ernst} F. Ernst, Phys. Rev. 167 (1968) 1175.

\bibitem{preseg} A. Pressley and G. Segal, Loop groups, Oxford (1986) Clarendon Press.

\bibitem{kac} V.G. Kac, ``{\it Infinite dimensional Lie algebras.}, Cambridge
Univ. Press, 1990.

\bibitem{babono} O. Babelon and L. Bonora, Phys. Lett. B267 (1991) 71. 

\bibitem{ueno} K. Ueno and Y. Nakamura, Phys. Lett. 109B (1982) 273.

\bibitem{babe} O. Babelon and D. Bernard, Commun. Math. Phys. 149 (1992) 279.

\bibitem{babe2} O. Babelon and D. Bernard, Int. J. Mod. Phys. A8 (1993) 507.

%
\end{thebibliography}
\end{document}